\documentclass{JHEP3}    
\usepackage{epsfig}

\newcommand{\eref}[1]{(\ref{#1})}

\newcommand{\fref}[1]{Figure~\ref{#1}}
\newcommand{\cref}[1]{Chapter~\ref{#1}}
\newcommand{\beq}{\begin{equation}}
\newcommand{\eeq}{\end{equation}}
\newcommand{\ba}{\begin{array}}
\newcommand{\ea}{\end{array}}
\newcommand{\bcenter}{\begin{center}}
\newcommand{\ecenter}{\end{center}}

\def\IB{\relax\hbox{$\inbar\kern-.3em{\rm B}$}}
\def\IC{\relax\hbox{$\inbar\kern-.3em{\rm C}$}}
\def\ID{\relax\hbox{$\inbar\kern-.3em{\rm D}$}}
\def\IE{\relax\hbox{$\inbar\kern-.3em{\rm E}$}}
\def\IF{\relax\hbox{$\inbar\kern-.3em{\rm F}$}}
\def\IG{\relax\hbox{$\inbar\kern-.3em{\rm G}$}}
\def\IGa{\relax\hbox{${\rm I}\kern-.18em\Gamma$}}
\def\IH{\relax{\rm I\kern-.18em H}}
\def\IK{\relax{\rm I\kern-.18em K}}
\def\IL{\relax{\rm I\kern-.18em L}}
\def\IP{\relax{\rm I\kern-.18em P}}
\def\IR{\relax{\rm I\kern-.18em R}}
\def\IZ{\relax\ifmmode\mathchoice
{\hbox{\cmss Z\kern-.4em Z}}{\hbox{\cmss Z\kern-.4em Z}}
{\lower.9pt\hbox{\cmsss Z\kern-.4em Z}}
{\lower1.2pt\hbox{\cmsss Z\kern-.4em Z}}\else{\cmss Z\kern-.4em Z}\fi}
\def\II{\relax{\rm I\kern-.18em I}}

\def\sCC{{\kern 0.27em\vrule height1.45ex width0.03em depth0em
          \kern-0.30em\rm C}}
\def\C{{\mathchoice
  {\sCC}
  {\sCC}
  {\kern 0.225em \vrule height1.05ex width0.025em depth0em \kern-0.25em \rm C}
  {\kern 0.180em \vrule height0.78ex width0.02em depth0em \kern-0.2em \rm C}
        }}
\def\sHH{{\rm I\kern-.16em{}H}}
\def\H{{\mathchoice
  {\sHH}
  {\sHH}
  {\rm I\kern-.13em{}H}
  {\rm I\kern-.13em{}H} }}
\def\sNN{{\rm I\kern-.16em{}N}}
\def\N{{\mathchoice
  {\sNN}
  {\sNN}
  {\rm I\kern-.12em{}N}
  {\rm I\kern-.10em{}N} }}
\def\sPP{{\rm I\kern-.16em{}P}}
\def\P{{\mathchoice
  {\sPP}
  {\sPP}
  {\rm I\kern-.12em{}P}
  {\rm I\kern-.10em{}P} }}
\def\sQQ{{\kern 0.27em \vrule height1.45ex width0.03em depth0em
          \kern-0.30em \rm Q}}
\def\Q{{\mathchoice
        {\sQQ}
        {\sQQ}
  {\kern 0.225em \vrule height1.05ex width0.025em depth0em \kern-0.25em \rm Q}
  {\kern 0.180em \vrule height0.78ex width0.020em depth0em \kern-0.20em \rm Q}
        }}
\def\sRR{{\rm I\kern-0.16em{}R}}
\def\R{{\mathchoice
  {\sRR}
  {\sRR}
  {\rm I\kern-0.12em{}R}
  {\rm I\kern-0.10em{}R} }}
\def\sZZ{{\rm Z\kern-0.32em{}Z}}
\def\Z{{\mathchoice
  {\sZZ}
  {\sZZ} 
  {\rm Z\kern-0.3em{}Z}     
  {\rm Z\kern-0.25em{}Z} }}  
\def\ZZZ{{\rm Z\kern-0.24em{}Z}}
\def\sII{{\rm I\kern-0.16em{}I}}
\def\I{{\mathchoice
  {\sII}
  {\sII}
  {\rm I\kern-0.12em{}I}
  {\rm I\kern-0.10em{}I} }}

\def\inbar{\,\vrule height1.5ex width.4pt depth0pt}
\font\cmss=cmss10 \font\cmsss=cmss10 at 7pt

\def\smiley{\hbox{\large$\bigcirc$\hspace{-0.80em}\raise.2ex
\hbox{$\cdot\cdot$}\kern-.61em\lower.2ex\hbox{\scriptsize$\smile$}}\ }
\def\frowny{\hbox{\large$\bigcirc$\hspace{-0.80em}\raise.2ex
\hbox{$\cdot\cdot$}\kern-.635em\lower.2ex\hbox{\scriptsize$\frown$}}\ }

\def\I{{\rlap{1} \hskip 1.6pt \hbox{1}}}

\makeatletter
\let\hangafter\@hangfrom
\makeatother

\newcommand{\be}{\begin{equation}}
\newcommand{\ee}{\end{equation}}
\newcommand{\bea}{\begin{eqnarray}}
\newcommand{\eea}{\end{eqnarray}}
\newcommand{\bean}{\begin{eqnarray*}}
\newcommand{\eean}{\end{eqnarray*}}

\newcommand{\beqa}{\begin{eqnarray}}
\newcommand{\eeqa}{\end{eqnarray}}

\preprint{MIT-CTP-3627\\ IFT-UAM/CSIC-05-20\\ 
CERN-TH-PH/2005-055\\ hep-th/0505040}

\title{Fractional Branes and Dynamical Supersymmetry Breaking}

\author{Sebasti\'an Franco$^\dagger$, Amihay Hanany$^\dagger$, Fouad Saad 
$^\ddagger$, Angel M. Uranga$^\ddagger$
\footnote{
Research supported in part by the CTP and the LNS
of MIT and the U.S. Department of Energy under cooperative agreement
$\#$DE-FC02-94ER40818, and the CICYT, Spain. A. H. is also supported by 
BSF American--Israeli Bi--National Science Foundation and a DOE OJI award.}
\\
~\\
$^\dagger$Center for Theoretical Physics,
\\ Massachusetts Institute of Technology,\\
Cambridge, MA 02139, USA.\\
\email{sfranco, hanany@mit.edu} \\
$^\#$Instituto de F\'{\i}sica Te\'orica, Facultad de Ciencias, C- XVI \\
Universidad Aut\'onoma de Madrid, 28049 Madrid, Spain\\
and \\
TH Division, CERN, CH-1211 Geneve 23, Switzerland\\
\email{fouad.saad, angel.uranga@uam.es, angel.uranga@cern.ch}
}

\abstract{
We study the dynamics of fractional branes at toric singularities, including cones 
over del Pezzo surfaces and the recently constructed $Y^{p,q}$ theories. We find 
that generically the field theories on such fractional branes show dynamical 
supersymmetry breaking, due to the appearance of non-perturbative superpotentials. 
In special cases, one recovers the known cases of supersymmetric infrared 
behaviors, associated to SYM confinement (mapped to complex deformations of the 
dual geometries, in the gauge/string correspondence sense) or $N=2$ fractional 
branes. In the supersymmetry breaking cases, when the dynamics of closed string 
moduli at the singularity is included, the theories show a runaway behavior (involving 
moduli such as FI terms or equivalently dibaryonic operators), rather than stable non-supersymmetric 
minima. We comment on the implications of this gauge theory behavior for the
infrared smoothing of the dual warped throat solutions with 3-form fluxes, 
describing duality cascades ending in such field theories. We finally provide a 
description of the different fractional branes in the recently introduced brane 
tiling configurations.
}
\begin{document}

\section{Introduction}

D-branes at singularities provide a useful arena to study and test the gauge/string 
correspondence, in situations with reduced (super)symmetry. When D3-branes are 
located at a conical singularity, they lead to 
conformal field theories, described as quiver gauge theories, whose dual is 
provided by Type IIB theory on $AdS_5\times X_5$ where $X_5$ is the base of the real cone
given by the singularity. 

A simple way to break conformal invariance is to introduce fractional branes at the 
singularity. Physically they correspond to D-branes wrapped on cycles collapsed at 
the singularity, consistently with cancellation of (local) RR tadpoles. At the level 
of the quiver, fractional branes correspond to rank assignments for nodes in the 
quiver, which are consistent with cancellation of non-abelian anomalies.

In the presence of a large number of D3-branes, a small amount of fractional branes 
leads to a controlled breaking of conformal invariance, yielding field theories with 
tractable supergravity duals. In many cases, the field theory RG flow is known to 
lead to a cascade of Seiberg dualities in which the effective number of D3-branes 
decreases, while that of fractional branes remains constant 
\cite{Klebanov:2000hb,Franco:2004jz,Ejaz:2004tr,Franco:2005fd}. Hence the infrared dynamics is 
dominated by the field theory on the fractional branes, in the 
absence of D3-branes. Hence, the exact dynamics (including non-perturbative effects) 
of the field theories on fractional branes is an important question. In fact, it is 
an interesting question even independently of whether such field theories lie at the 
end of duality cascades or not.

A systematic study of fractional branes, and the dynamics they trigger, for large 
classes of toric singularities was initiated in \cite{Franco:2005fd}. There, 
different tools were provided to identify fractional branes leading to infrared 
confinement, described in the supergravity dual as complex deformations of the 
conical geometry. A precise description of the complex deformation is obtained by 
using the web diagram for the singularity, dual to its toric diagram \cite{Aharony:1997bh,Leung:1997tw}, in terms of 
removal of a subweb in equilibrium. The gauge 
dynamics, and its dual supergravity description, is very similar to that in 
\cite{Klebanov:2000hb}. The cases covered by this analysis include a subset of 
the throat solutions studied in \cite{Franco:2004jz} for cones over del Pezzo 
surfaces, and the complex deformation provides the appropriate smoothing of the 
infrared singularity of these throats.

On the other hand, other choices of fractional branes do not trigger complex 
deformations, since they do not correspond to removal of subwebs in equilibrium. 
We will see that this applies to most of the fractional branes employed in the 
throat solutions for cones over del Pezzo surfaces \cite{Franco:2004jz} and cones 
over the $Y^{p,q}$ manifolds \cite{Ejaz:2004tr}. This nicely agrees with 
the fact that the field theories on the corresponding fractional branes do not 
show simple confinement as in the previous family. The infrared dynamics of these 
gauge field theories is thus an important hint in understanding the 
properties of these supergravity solutions, and in particular the smoothing of their
infrared naked singularities.

In this paper we address the general question of analyzing the dynamics of the 
gauge field theories on fractional branes. The results are very interesting. We show 
that generically the field theories show Dynamical Supersymmetry Breaking (DSB). The 
origin of DSB is an incompatibility between the fact that the branes are fractional 
(so that there are no classical flat directions and the classical F- and D- scalar 
potential forces all vevs to vanish) and the dynamical generation of a 
non-perturbative Affleck-Dine-Seiberg (ADS) superpotential (which pushes the meson 
vevs to infinity, namely repels the branes from the origin).

More precisely, once the coupling of closed string moduli to the $U(1)$ factors as 
Fayet-Illiopoulos (FI) terms is taken into account (equivalently, once $U(1)$'s are 
made massive by $B\wedge F$ couplings and disappear, thus allowing for the 
existence of dibaryonic operators), the theories show a runaway behavior in those 
directions. Hence the complete dynamics strictly correspond to a runaway situation, 
rather than to a stable non-supersymmetric minimum. This is similar to the 
discussion in an orientifold model in \cite{Lykken:1998ec}. Concerning the 
supergravity duals, the gauge analysis suggests that it may be possible to construct 
a geometric smoothing of the naked singularities in the warped throat solutions, 
but it should be non-supersymmetric and presumably non-stable. Namely there does not 
seem to exist a stable smooth geometry whose UV asymptotics correspond to the 
warped throats solutions with 3-form fluxes (for choices of fluxes corresponding to 
DSB fractional branes).

In this paper we perform an exhaustive discussion of different kinds of fractional 
branes and their dynamics in several families of examples, including the suspended 
pinch point (SPP) singularity, cones over del Pezzo surfaces, and cones over the 
$Y^{p,q}$ manifolds. Many other examples, recently available using the techniques in 
e.g. \cite{Hanany:2005hq} can be worked out similarly. In particular the use of 
brane dimers and brane tilings \cite{Hanany:2005ve,Franco:2005rj} may allow to 
sharpen early discussions of dynamical susy breaking using brane configurations 
\cite{Hanany:1997tb}. In fact, we take some steps in this direction by providing a 
detailed description in terms of brane tilings of fractional branes with 
supersymmetric infrared dynamics.

The paper is organized as follows. In section \ref{section_classification_fractional}
we classify fractional branes in three different classes, according to the infrared dynamics they produce. They are named i) deformation branes, triggering complex 
deformations in the supergravity dual, ii) $N=2$ branes, leading to $N=2$ dynamics 
along its flat direction, and iii) DSB branes, whose dynamics breaks supersymmetry. 
In section \ref{dp1} we study the simplest example of a DSB fractional brane, 
arising for the cone over $dP_1$. We study the gauge theory on fractional branes 
from different viewpoints, and describe the breakdown of supersymmetry due to 
non-perturbative effects. We discuss the role of closed string 
moduli at the singularity (equivalently of di-baryonic operators on the gauge 
theory), arguing that the behavior of the theory is a runaway one, and does not really have a 
stable non-supersymmetric minimum. 

In sections \ref{sppexample} and \ref{higherdpn} we carry out a similar analysis for 
other theories, including the SPP singularity and some examples of cones over del 
Pezzo surfaces. We show that combinations of different deformation and/or $N=2$ 
fractional branes, which individually lead to supersymmetric infrared behavior, may 
lead to breakdown of supersymmetry when combined together, and present several 
examples of this kind. In section \ref{ypqfamily} we center on the infinite family 
of $Y^{p,q}$ theories, and show that the fractional brane leads to breakdown of 
supersymmetry in all these theories. This should have important implications for the 
infrared smoothing of the warped throat solutions in \cite{Ejaz:2004tr}, on which 
we comment.

In section \ref{brane_tiling} we provide a description of fractional branes in terms 
of the brane dimers and brane tiling constructions recently introduced in \cite{Hanany:2005ve,Franco:2005rj}. This 
provides a simple procedure for describing the infrared dynamics for supersymmetric 
branes, like the effect of infrared confinement for deformation branes. We also 
use these techniques to describe a subset of the baryonic $U(1)$ global symmetries of the gauge theory.

Finally, in section \ref{conclusions} we offer our final comments. In appendix
\ref{obstruction} we review the mathematical description of complex deformations for 
toric singularities, and show that the cones over $Y^{p,q}$ geometries ($q\neq 0$) 
have first order deformations, but they are obstructed at second order.

\bigskip
{\bf Note:} While this paper was ready for submission, we became aware of the work of \cite{Berenstein:2005xa} which 
has some overlap with this paper.

\medskip

\section{Classes of fractional branes}

\label{section_classification_fractional}

Fractional branes correspond to higher-dimensional D-branes wrapped over collapsed 
cycles in the 
singularity. From the point of view of the gauge theory, they correspond to anomaly 
free rank assignments for the quiver nodes. Hence, they are associated to vectors in 
the kernel of the antisymmetric intersection matrix defining the quiver. 
They can be classified into three groups, according to the different IR behaviors that
they trigger (see Section \ref{brane_tiling} for a description of the different 
fractional branes using brane tilings).

\bigskip

\noindent{\bf $N=2$ fractional branes:} The field theory on these fractional branes 
has flat directions along which the dynamics generically reduces to an $N=2$ theory. The 
simplest examples are provided by fractional branes whose quiver (i.e. the quiver 
in the absence of any other type of fractional branes or probe D3-branes) 
corresponds to a closed loop of arrows passing through all nodes, with the 
corresponding gauge invariant polynomial not appearing in the superpotential. The 
vev for this operator is F-flat, and parametrizes a one-dimensional moduli space, 
along  which the dynamics has an accidental $N=2$ supersymmetry (8 supercharges) 
and in the simplest case is described by an $N=2$ SYM theory. 

Geometrically, these fractional branes appear 
for non-isolated singularities, which have (complex) curves of singularities 
passing through the origin. The fractional branes wrap the 2-cycles collapsed at the 
singularity, which exist at any point in the curve. For toric geometries, the 
singularity on the curve is always of $\IC^2/\IZ_N$ type. The curves of 
singularities are associated to the existence of points on the boundary of the toric 
polygon, or equivalently to parallel semi-infinite legs in the dual web diagram. 

These fractional branes lead to supersymmetric infrared dynamics, which can be 
described in terms of the corresponding Seiberg-Witten curve. In the dual 
supergravity, they lead to enhan\c{c}on like backgrounds 
\cite{Johnson:1999qt,Bertolini:2000dk,Gauntlett:2001ps,Petrini:2001fk}.

\bigskip

\noindent{\bf `Deformation' fractional branes:} The corresponding 
quiver field theory corresponds to either a set of decoupled nodes (only gauge groups and 
no bifundamental fields), or to a closed 
loop of arrows with the corresponding gauge invariant appearing in the 
superpotential. The theory thus does not have flat directions. The dynamics of the 
field theory is (possibly partial) confinement. In the supergravity dual, the 
geometry undergoes a complex deformation that smoothes (possibly partially) the 
conical geometry. In terms of the web diagram, the complex deformation corresponds 
to the removal from the diagram of a subweb in equilibrium. These fractional branes, 
and the corresponding complex deformation have been extensively studied in many 
examples in \cite{Franco:2005fd}. 

\bigskip

\noindent{\bf DSB fractional branes:} Any other anomaly free rank assignment in the 
quiver seems 
to lead to dynamical supersymmetry breaking. The prototypical field theory for such 
fractional brane corresponds to a set of nodes of generically different ranks, with 
bifundamental matter. The non-perturbative dynamics typically contains a 
contribution from a node 
generating a non-perturbative ADS superpotential. In cases with classical flat 
directions, they are lifted in a runaway fashion by this superpotential. In cases 
without flat directions, DSB arises from an incompatibility between the ADS 
superpotential and the classical potential forcing all vevs to vanish (as mentioned 
above, one recovers runaway behavior when Fayet-Illiopoulos (FI) terms are 
considered dynamical, or equivalently when one eliminates the $U(1)$ factors and 
allows for dibaryonic operators).

An important point is that the generic fractional brane case falls in this class. 
More concretely, generically the combination of an $N=2$ fractional brane with a 
deformation fractional brane is a DSB combination. Also, in general the 
combination of two deformation fractional branes, for different and 
incompatible deformations, is also a DSB fractional brane. 

Our main interest in this paper is the infrared dynamics of these fractional branes. 
We hope that the gauge theory results provide useful information to describe the IR 
completion of the supergravity dual throat solutions with fluxes.

\section{The $dP_1$ case}
\label{dp1}

\begin{figure}[ht]
  \epsfxsize = 3cm
  \centerline{\epsfbox{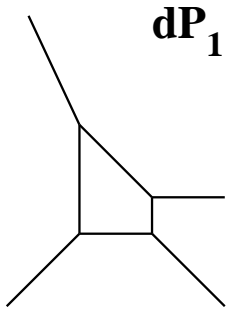}}
  \caption{\small Web diagram for the cone over $dP_1$}
  \label{webdp1}
\end{figure}

The $dP_1$ theory provides the simplest example of a duality cascade with 
fractional branes, where the infrared behavior is not described by confinement/complex 
deformation (or by $N=2$ like dynamics). Namely, from the web diagram in figure 
\ref{webdp1}, there is no possibility of splitting a subweb in equilibrium
\cite{Franco:2005fd}, hence there are no complex deformations of this geometry. 
Naively this seems in contradiction with the recent results in 
\cite{Burrington:2005zd}, where a first order deformation of the throat solution 
with fluxes for the $Y^{p,q}$ geometries (and hence for the cone over $dP_1$ 
which corresponds to the $Y^{2,1}$ theory) was explicitly constructed. However, as 
described in \cite{Morrison:1996xf} (see also appendix \ref{obstruction}), the cone 
over $dP_1$ precisely has an obstructed deformation, namely a first order 
deformation, which is obstructed at second order. Hence this result nicely 
reconciles both statements, and confirms that a smoothing of the infrared 
singularity by a complex deformation is not possible.

In this section we study the $dP_1$ gauge theory in detail, describing our proposal 
for the IR behavior.

\subsection{Quiver theory and UV cascade}

Let us consider the cone over $dP_1$ geometry. The quiver field theory on 
D3-branes at this singularity has been constructed in
\cite{Feng:2000mi}. Diverse aspects of this theory have been studied in
\cite{Feng:2001xr,Feng:2001bn,Feng:2002zw,Feng:2002fv,Franco:2002ae,Intriligator:2003wr,Bertolini:2004xf,Benvenuti:2004dy,Franco:2005fd} 

\begin{figure}[ht]
  \epsfxsize = 3cm
  \centerline{\epsfbox{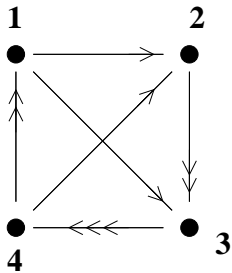}}
  \caption{\small Quiver diagram for the $dP_1$ theory}
  \label{quiverdp1}
\end{figure}

Out of the different Seiberg dual theories corresponding to this geometry, we
focus on the phase with quiver diagram shown in \fref{quiverdp1} and 
superpotential
\beq
W=
 \epsilon_{\alpha \beta} X^{\alpha}_{23} X^{\beta}_{34} X_{42}  
+\epsilon_{\alpha \beta} X^{\alpha}_{34} X^{\beta}_{41} X_{13}
-\epsilon_{\alpha \beta} X_{12} X^{\alpha}_{23} X^3_{34} X^{\beta}_{41} 
\label{W_dP1}
\eeq
where $\alpha,\beta$ take values in $1,2$ and label a doublet representation of 
$SU(2)$. Also, subindices indicate the bifundamental representation under the 
corresponding nodes. The web diagram is shown in figure \ref{webdp1}.

\begin{figure}[ht]
  \epsfxsize = 3cm
  \centerline{\epsfbox{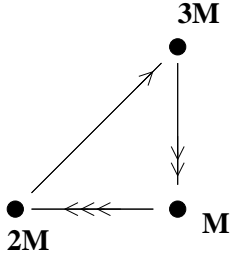}}
  \caption{\small The theory at the end of the duality cascade triggered by $M$
fractional branes. Here labels indicate ranks for the node gauge factors.}
  \label{quiverdp1end}
\end{figure}

There is only one kind of fractional brane\footnote{For a web diagram with $n$ external legs the number of fractional branes is $n-3$. See \cite{Hanany:2005hq} for a discussion on this point and a comparison to the parameters of the corresponding 5-dimensional theory \cite{Aharony:1997bh}. A discussion of this point from a geometric viewpoint was recently done in \cite{Martelli:2005tp}.}, corresponding to the rank vector $(0,3,1,2)$. As we explained above, this can be determined by looking at anomaly free rank assignments for the gauge groups. Starting with ranks $(N,N+3M,N+M,N+2M)$, this
fractional brane triggers a duality cascade proposed in \cite{Franco:2004jz}
and verified in \cite{Ejaz:2004tr,Franco:2005fd}. Along the cascade, the effective 
number of D3-branes decreases, while the number of fractional branes remains 
constant. Hence, for suitable UV choice of the number of D3-branes $N$, the infrared limit
of the cascade is expected to be described by the theory in figure \ref{quiverdp1end}, 
in which the smallest rank node has reached zero rank and disappeared from the quiver.

Notice that this gauge theory does not correspond to the quiver of a deformation 
brane, as described in section \ref{section_classification_fractional}. For 
instance, it does not correspond to a set of decoupled $SU(M)$ SYM 
theories without matter, as it happens in the absence of D3-branes in the theories 
studied in \cite{Franco:2005fd}. The quiver in figure \ref{quiverdp1end} should be 
contrasted with what happens for example in the well studied example of the 
conifold, where there are no bifundamental fields unless D3-branes are included. 
Hence, we expect a behavior which is qualitatively different from that of the 
conifold \cite{Klebanov:2000hb}. Indeed, this is also 
supported by the geometric side, since the web diagram in figure \ref{webdp1} does 
not admit a recombination of external legs into a subweb in equilibrium 
\cite{Franco:2005fd}. Namely, the cone over $dP_1$ does not admit a complex 
deformation or extremal transition in which 2- and 4-cycles disappear and 3-cycles grow 
\cite{Morrison:1996xf}.

It is an interesting question to find the field theory dynamics which dominates the infrared limit of 
this cascade, and its corresponding gravity dual. In the coming sections we carry out a 
simple field theory analysis to argue that the answer is dynamical supersymmetry breaking.

\subsection{Dynamical supersymmetry breaking in $dP_1$}
\label{dsbdp1}

\subsubsection{Field theory analysis}
\label{fieldthdp1}

The dynamics of the infrared limit of the cascade is controlled by the quiver theory of figure \ref{quiverdp1end}. It corresponds to considering $M$ fractional D-branes, without 
any D3-branes, $N=0$. It leads to a theory with gauge group $SU(3M)_2\times SU(M)_3\times 
SU(2M)_4$, with fields $X_{42}$, $X_{23}=X_{23}^1$, $Y_{23}=X_{23}^2$, $X_{34}=X_{34}^1$, $Y_{34}=X_{34}^2$, $Z_{34}=X_{34}^3$  
(were we have simplified notation with respect to that of equation (\ref{W_dP1}).), and superpotential
\beqa
W= X_{42}X_{23}Y_{34} - X_{42}Y_{23}X_{34}.
\eeqa

There are several ways to support the idea of the onset of dynamical supersymmetry 
breaking in this theory, by using several standard criteria (see 
\cite{Shadmi:1999jy} for a very complete introduction to DSB). One of the simplest ways
is as follows: Consider a theory without classical flat directions, and such that the 
classical D- and F-term constraints force all vevs of the theory to vanish. In 
such a situation, if one of the gauge factors of the theory has $N_f<N_c$, then the 
non-perturbative Affleck-Dine-Seiberg superpotential for its mesons diverges at 
the origin, and pushes the corresponding vevs towards infinity. The theory breaks 
supersymmetry due to the impossibility to satisfy all F- and D-term constraints, coming from classical and quantum contributions. The combination of the classical and the non-perturbative superpotential lead to a scalar potential with a minimum at non-zero energy. We will apply this idea to various gauge theories along this paper. We now proceed to a detailed application to the $dP_1$ case.

Furthermore, based on the physical interpretation of fractional branes, it is easy to realize that 
the above theory does not have flat directions (since they would correspond to 
removal of the branes out of the singularity, which is not possible for fractional 
branes). This can also be directly recovered from the field theory analysis, by 
looking for D- and F-flat directions. However, a crucial issue in getting the 
correct result is the following. The string theory construction leads to a gauge group 
$U(3M)\times U(M)\times U(2M)$. The three $U(1)$ factors in this gauge group have 
$B\wedge F$ couplings to 2-forms which are localized at the singularity (these arise 
from reduction of the RR 6- and 4-form on the 4- and the two 2-cycles on the cone over 
$dP_1$). These couplings (which are crucial in the Green-Schwarz cancellation of 
mixed anomalies, as in \cite{Ibanez:1998qp}) make the $U(1)$'s massive, so that 
they are not present at low energies. On the other hand, the D-term constraints 
with respect to these $U(1)$'s remain, and have to be taken into account in order to derive 
the correct moduli space. Notice that this is implicit in the familiar statement 
(implied by supersymmetry) that the NSNS partners of the above RR fields couple to 
the D-branes as Fayet-Illiopoulos terms 
\cite{Douglas:1996sw,Douglas:1997de,Morrison:1998cs}.

We thus parametrize D-flat directions by operators invariant under the  $SU(3M)\times 
SU(M)\times SU(2M)$ gauge symmetry. There are 6 such operators,
\beqa
X_{42}X_{23}X_{34}
,\quad  X_{42}X_{23}Y_{34}
,\quad 
X_{42}X_{23}Z_{34},
\nonumber \\ 
X_{42}Y_{23}X_{34}
,\quad  X_{42}Y_{23}Y_{34}
,\quad 
X_{42}Y_{23}Z_{34}.
\eeqa

In order to impose F-flatness, we use e.g. the equations of motion
\beqa
\frac{\partial W}{\partial Y_{34}}= X_{42}X_{23}=0\quad , \quad
\frac{\partial W}{\partial X_{34}}= X_{42}Y_{23}=0
\eeqa
so that all operators are forced to vanish, and the origin is the only supersymmetric 
point. The classical superpotential thus lifts all flat directions.

It is now easy to argue that this theory breaks supersymmetry. Consider the regime 
where the $SU(3M)$ gauge factor dominates the dynamics. Since $SU(3M)$ has $2M$ flavors, 
we have $N_f<N_c$ for this theory, and it generates a non-perturbative Affleck-Dine-Seiberg  
superpotential which pushes the vevs for the $SU(3M)$ mesons $X_{42}X_{23}$ and 
$X_{42}Y_{23}$ away from zero. Combining this with the classical superpotential, we conclude that supersymmetry is broken. A more detailed 
analysis is presented below, in a slightly different limit.

An independent argument for DSB in this theory follows from the following  
alternative criterion. In a theory with no classical flat directions and with some 
spontaneously broken global symmetry, supersymmetry breaking occurs (see e.g. 
\cite{Shadmi:1999jy}). The argument is that the complex scalar in the Goldstone 
supermultiplet would parametrize a non-compact flat direction, which would reach 
the semiclassical regime, in contradiction with the absence of classical flat 
directions. With supersymmetry breaking, the Goldstone boson still parametrizes a 
compact flat direction, but the non-compact direction associated to its partner is 
lifted. In our case, the theory originally has a global $SU(2)$ symmetry, under which the  
$SU(3M)$ mesons $X_{42}X_{23}$ and $X_{42}Y_{23}$ transform in the spin-1/2 
representation. The global $SU(2)$ is thus spontaneously broken by the meson vevs 
triggered by the ADS superpotential. Hence the criterion for DSB is satisfied.

The physical realization of the gauge field theory in terms of fractional D-branes 
makes also clear that there should exist a non-supersymmetric minimum at finite 
distance in field space, since the scalar potential grows both for large and 
small vevs. By considering the regime where the ADS superpotential is dominant over 
the classical one, one could make the minimum lie in the semiclassical region, so 
that a perturbative analysis would be possible. However, it is not clear that the 
fractional brane system has a tunable parameter that allows to consider this limit.

In any event, some qualitative features of the remaining theory at the minimum can 
be suggested. By taking the most symmetric choice of $SU(3M)$ meson vevs ${\cal 
M}={\bf 1}$, the gauge symmetry $SU(2M) \times SU(M)$ is broken to just $SU(M)$, 
the diagonal subgroup of $SU(M)\times SU(M)\times SU(M)\subset SU(2M)\times SU(M)$. 
In addition, the superpotential makes the mesons massive together with the $X_{34}$ 
and $Y_{34}$ fields. Hence at the minimum we have an $SU(M)$ theory with some 
adjoint matter (coming e.g. from the $Z_{34}$ fields). A more detailed description 
of the minimum is also possible in other regimes, e.g. when the $SU(2M)$ dynamics dominates (see below).

Notice that the above discussion is similar to the original analysis of the $(3,2)$ 
model in \cite{Affleck:1984xz}. In fact, for a single fractional brane $M=1$, the theory 
is very reminiscent of the $(3,2)$ model. The main difference is that it contains 
some additional doublets, and that the flat directions are removed by a combination 
of F-terms and the additional $U(1)$ D-flatness conditions. Also, a similar model 
was studied in \cite{Hanany:1997tb}.

\medskip

It is interesting to point out that the D-brane realization of this system provides 
a representation of the non-perturbative effect leading to the ADS superpotential.
The non-perturbative effect is generated by euclidean D1-branes wrapped on the 
2-cycle associated to the fractional brane. Such D-brane instanton preserves half of 
the four supersymmetries of the theory, thus has two fermion zero modes and can 
contribute to the superpotential of the theory. More precisely, these euclidean 
D-branes are fractional (where here `fractional' has the same meaning as for 
the fractional euclidean D-branes generating the gaugino condensate in D-brane 
realizations of $N=1$ SYM). Using dualities, this string theory 
interpretation can be related to a similar effect in \cite{Aharony:1997ju}.

\subsubsection{Analysis in a different regime}

The physics of the model can be analyzed also in other regimes, e.g. when
the $SU(2M)$ dynamics dominates, as follows. Consider the above $SU(3M)_2\times 
SU(M)_3\times SU(2M)_4$, with weakly gauged $SU(3M)\times SU(M)$, and dynamics 
dominated by the $SU(2M)$ factor. We can analyze the resulting dynamics by 
replacing it by its Seiberg dual. The gauge group of the resulting theory is 
$SU(3M)_2\times SU(M)_3\times SU(M)_4$. The fields include the original $SU(2M)$ 
singlets $X_{23}$, $Y_{23}$, dual quarks $X_{24}$, $X_{43}$, $Y_{43}$, $Z_{43}$ and 
mesons $M_{32}(=X_{34}X_{42})$, $N_{32}(=Y_{34}X_{42})$, $P_{32}(=Z_{34}X_{42})$. 
The superpotential is
\beqa
W= N_{32}X_{23}-M_{32}Y_{23}+ X_{24} \left( X_{43} M_{32}+ Y_{43} N_{32} + 
Z_{43} P_{32} \right)
\eeqa
The first two terms give masses to the corresponding fields. Integrating them out, 
we are left with 
a theory $SU(3M)_2\times SU(M)_3\times SU(M)_4$, fields $X_{24}$, $X_{43}$, $Y_{43}$, 
$Z_{43}$, $P_{32}$, and $W= X_{24}Z_{43} P_{32}$. The $SU(2M)$ dynamics thus 
preserves supersymmetry. However, dynamical supersymmetry breaking is recovered when we 
consider the $SU(3M)$ dynamics, which generates an ADS superpotential in a theory 
that does not have flat directions. Notice that it would seem that diagonal vevs 
e.g. for fields $X_{43}$, $Y_{43}$ are flat directions, but as discussed above they 
do not satisfy the additional $U(1)$ D-flatness conditions.

The resulting physics is most easily analyzed in the case $M=1$. There the $SU(3)$ 
theory confines, leaving the meson $M_{34}=P_{32}X_{24}$ as the only degree of 
freedom. The gauge group is trivial and we have a superpotential
\beqa
W= M_{34}Z_{43} + 2 \left( \frac{\Lambda^{8}}{M_{34}} \right)^{1/2}
\eeqa
The theory breaks supersymmetry since the F-term constraints cannot be satisfied.
There is a non-supersymmetric minimum, i.e. a minimum with non-zero energy, whose 
existence can be suggested as follows. The F-term scalar potential is
\beqa
V_F\, =\, |M_{34}|^2\, +\, \left|\, Z_{43}\, +\, \Lambda^4 M_{34}^{-3/2}\, \right|^2
\eeqa
This contribution alone would clearly generate a runaway behavior towards $M_{34}\to 0$ and $Z_{43}=-\Lambda^4 
M_{34}^{-3/2}\to \infty$, with a minimum at infinity. On the other hand, $Z_{43}$ also 
appears in the additional $U(1)$ D-term potential, of the form $V_D= |Z_{43}|^4$, 
which grows for large $Z_{43}$ vevs for fixed FI parameter, becoming a `barrier' that prevents
$Z_{43}$ for running away. The combination of these 
two contributions establishes the existence of the 
non-supersymmetric minimum for fixed FI terms, but a more quantitative analysis is unreliable since 
there is seemingly no tunable parameter which allows to make the minimum lie in the 
semiclassical large vev region. In the next section we discuss the effect 
of including the dynamics of FI terms in the analysis.

\subsubsection{Dynamical FI terms}

\label{section_dynamical_FI}

In this section we discuss an important fact that was not incorporated in the above determination of the non-trivial 
minimum. Namely, closed string fields at the 
singularity are dynamical, and couple as FI terms. Taking that into account, the 
relevant part of the D-term potential should be written as $V_D= (|Z_{43}|^2-\xi)^2$. 
This shows that one can afford to take large vevs for $Z_{43}$ (hence making the 
F-terms arbitrarily small) by simply allowing for a large FI $\xi$ (keeping the 
D-terms vanishing). Namely the system relaxes to minimization of its potential by 
dynamically allowing the closed string modes to blow up the singularity.

An equivalent description, which phrases the discussion purely in terms of the 
effective low energy field theory and is perhaps more appropriate for the 
context of the gauge/string correspondence, is as follows. As mentioned above, the 
coupling of the closed string RR fields, which are localized at the singularity, to the $U(1)$ factors 
generate a string scale mass for the latter. As a result, the $U(1)$ gauge multiplets disappear from the low energy physics. Hence, one should allow for (di)baryonic operators in the analysis 
of the low energy field theory. In particular the runaway direction described above 
is parametrized by the dibaryon $(Z_{43})^N$ associated to $Z_{43}$.
The relation between FI terms and dibaryons has been often discussed in the 
literature of brane configurations of the type considered in \cite{Hanany:1996ie}, 
see e.g. \cite{Elitzur:1997fh,Witten:1997sc}.

Notice that the discussion in terms of baryons allows to identify the runaway 
direction even in the original description of our field theory, in section 
\ref{fieldthdp1}. Namely it corresponds to the dibaryon 
\beqa
\epsilon_{i_1\ldots i_{2M}} \epsilon_{a_1\ldots 
a_M} \epsilon_{b_1\ldots b_M} (Z_{34})_{i_1a_1} \ldots (Z_{34})_{i_Ma_M}
(Z_{34})_{i_{M+1} b_1} \ldots (Z_{34})_{i_{2M} b_M}
\eeqa

Hence, we conclude that the full inclusion of localized closed string fields leads to 
runaway behavior in those directions. We want the reader to keep in mind that this situation will hold in subsequent examples, hence 
we skip the corresponding discussion and simply emphasize the different gauge 
theory dynamics, namely the open string sector. It would be interesting to have a more detailed understanding of this possible runaway behavior and its implications for the dual supergravity 
throat solutions. The natural suggestion is that a smoothing of the infrared naked 
singularities exists, which breaks supersymmetry, and which involves some kind of 
runaway instability. Hence the gauge theory suggests that there is no stable smooth 
solution with asymptotics corresponding to the warped throats with 3-form fluxes 
corresponding to DSB fractional branes.

\subsection{Additional D-brane probes}

In this section we comment on the interesting question of what happens in the 
presence of many fractional branes and a single or a few regular D3-brane probes. 
The classical moduli space of this theory is the moduli space of the additional 
D3-brane probe(s) (while the fractional ones remain stuck at the singular point). 
In the gravity dual, this translates to D3-branes able to probe the warped throat solution with Imaginary Self Dual (ISD) 3-form fluxes in \cite{Ejaz:2004tr}. This study hopefully provides some information on 
how fractional branes have modified its infrared naked singularity.

Consider the theory with ranks $(1,3M+1,M+1,2M+1)$. The gauge group is
$U(1)\times U(3M+1)\times U(M+1)\times U(2M+1)$ and following the 
discussion above we omit the $U(1)$ gauge groups since they become massive, thus leaving 3 groups in the product.
The bifundamental fields charged under the node with the $U(1)$ group and another of gauge groups should then be simply interpreted as fundamental flavors of the latter. The theory has a moduli space 
which corresponds to the position of the regular D3-brane in the cone over $dP_1$. 
Although a complete characterization is possible, it is sufficient for our purposes to focus on a particular one-dimensional flat direction. Consider the flat 
direction parametrized by the vev of the gauge invariant operator 
$Z_{34}X_{42}Y_{23}$. Clearly, the F-term constraints imply that other gauge 
invariant operators (e.g. $Z_{34}X_{41}Y_{13}$) get related vevs, but this will not 
concern us here. What is important is that it involves a vev for at least one 
component, denoted $\phi$ in what follows, of the matrix $X_{42}Y_{23}$, which 
is a meson of the gauge factor $SU(3M+1)$. 

Once non-perturbative dynamics is taken into account, it is clear that the theory with the additional D3-brane probe 
does not have a vacuum and in particular there is a runaway behavior for $\phi$. For 
instance, consider the regime where the $SU(3M+1)$ dynamics dominates. The 
$SU(3M+1)$ gauge group has $2M+2$ flavors, so it develops an ADS superpotential 
for the mesons, which pushes the determinant of the meson matrix
\beqa
{\cal M}= \pmatrix{X_{12}X_{23} & X_{12}Y_{23} \cr X_{42}X_{23} & X_{42}Y_{23} }
\eeqa
towards infinity. 

Without loss of generality we may perform gauge rotations to make the entry $\phi$ 
be the only non-vanishing one in its row and column. In that situation (and if we 
focus on the situation where the mesonic matrices $X_{42}X_{23}$, $X_{12}Y_{23}$, 
have vanishing entries in the corresponding row, column, respectively), $\phi$ 
splits off the mesonic matrix as a decoupled block. The superpotential then shows a 
manifest runaway behavior for $\phi$. Since $\phi$ is part of a D-flat direction, and 
is pushed to infinity by the F-term potential, we recover a runaway behavior along that 
classically flat direction. 

Physically, the D3-brane is repelled from the origin. Translating the gauge 
theory statement to the dynamics of a D3-brane probe of the warped throat solution 
with fluxes in \cite{Ejaz:2004tr}, the result shows that the D3-brane probe 
is repelled by the infrared resolution of the naked singularity. Interestingly, this 
is an independent argument that the mechanism to smooth out the singularity cannot 
be a complex deformation preserving the Calabi-Yau property and preserving the ISD 
character of the 3-form fluxes. There are general arguments that ISD 3-form fluxes 
on CY geometries lead to no forces on D3-brane probes 
\cite{Giddings:2001yu,Grana:2002nq,Camara:2003ku,Grana:2003ek}, in contrast with the 
repulsion felt by our D3-brane probe.

It would be nice to extract more information about the infrared nature of the 
solution from the gauge theory analysis. We leave this interesting question for 
future work.

\medskip

Finally, we would like to mention the possibility of combining different types of fractional branes 
to reach new physical situations and to illustrate that in general such  
combinations do not lead to a simple superposition of the behaviors associated to the individual branes. Consider for 
instance combining a fractional brane $(0,3,1,2)$ with a fractional brane 
$(3,0,2,1)$. Although each of them independently leads to DSB as described above, 
their combination adds up to the rank vector $(3,3,3,3)$ that corresponds to a 
set of D3-branes, which clearly preserves supersymmetry. In coming sections we will 
encounter other examples where combination of different fractional branes leads to 
results which are not the naive superposition of the individual behaviors.

\section{The SPP example}
\label{sppexample}

In this section we consider fractional branes in the suspended pinch point (SPP)
singularity, where a new effect takes place. In this theory, there exists two 
independent kinds of fractional branes which independently do not break 
supersymmetry. One kind leads to a duality cascade, confinement and complex 
deformation as studied in \cite{Franco:2005fd}, while the other belongs to an $N=2$ 
subsector, and leads to an enhan\c{c}on like behavior. However, as we will discuss, combinations of the 
two fractional branes may be incompatible and lead to runaway behavior (in this case even before 
considering localized closed string modes/baryonic directions).

\begin{figure}[ht]
  \epsfxsize = 3cm
  \centerline{\epsfbox{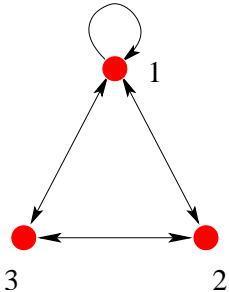}}
  \caption{\small The quiver for the SPP theory}
  \label{quiver_SPP}
\end{figure}

Consider the theory on D3-branes at an SPP singularity, defined by $xy=zw^2$. 
The theory, studied in \cite{Morrison:1998cs,Uranga:1998vf}, has a
quiver shown in figure \ref{quiver_SPP}

The superpotential is 
\beq
W=X_{21}X_{12}X_{23}X_{32}-X_{32}X_{23}X_{31}X_{13}+X_{13}X_{31}X_{11}-X_{12}X_{21}X_{11}
\eeq

\begin{figure}[ht]
  \epsfxsize = 8cm
  \centerline{\epsfbox{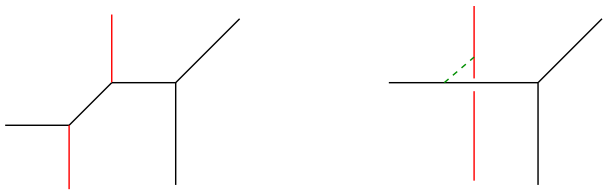}}
  \caption{\small Complex deformation for the SPP theory}
  \label{web_SPP}
\end{figure}

There are two independent fractional branes, and a basis for them is provided by the 
rank vectors $(1,0,0)$ and $(0,1,0)$. The physics of each independent fractional 
brane is well-known. The $(0,1,0)$ triggers the complex deformation studied in 
\cite{Franco:2005fd}, and shown in figure \ref{web_SPP}. The $(1,0,0)$ corresponds 
to a fractional brane of an $N=2$ subsector. Notice that this kind of fractional 
brane has a modulus, parametrized by 
the adjoint chiral multiplet, that corresponds to sliding the fractional brane 
along the curve of $A_1$ singularities of the geometry, parametrized by $z$ in 
$xy=zw^2$.

\begin{figure}[ht]
  \epsfxsize = 8cm
  \centerline{\epsfbox{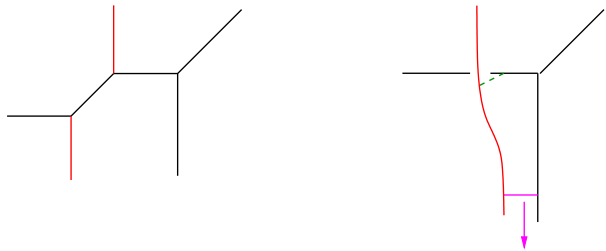}}
  \caption{\small Web picture of the incompatibility of complex deformation and 
$N=2$ fractional brane. The dashed segment represents the 3-cycle in the complex 
deformation, while the continuous segment represents the 2-cycle associated to the 
$N=2$ brane. The picture suggests a physical interpretation of the 
runaway behavior of ADS superpotentials in this case: The complex deformation 
increases the tension of the $N=2$ fractional brane, unless it escapes to infinity.}
  \label{ads}
\end{figure}

It is natural to consider what happens when both kinds of fractional branes are simultaneously
present. In the following, we consider the generic case where the numbers of 
fractional branes are different. Geometrically, there is an incompatibility between 
the branes, since the $(0,1,0)$ triggers a complex deformation that smoothes out the 
space, and hence also removes the curve of $A_1$ singularities, i.e. the collapsed 
2-cycle at the latter gets a finite size. In that situation, the $(1,0,0)$ branes 
which are wrapped over the 2-cycle get an additional tension, and break supersymmetry.
Supersymmetry would in principle be restored if the brane $(1,0,0)$ escapes to 
infinity along the curve of singularities as the complex deformation takes place. 
\fref{ads} gives a pictorial depiction of this situation. It would be interesting to understand whether this picture goes beyond being a nice intuitive representation and we can associate to it a more quantitative geometric meaning. The discussion is similar to that in 
\cite{Berenstein:2003fx}.

We may therefore expect a runaway behavior in this system. In order to verify this 
in detail, we take $M$ branes of type $(1,0,0)$ and $P$ branes of type $(0,1,0)$, 
and consider $P\gg M$. The dynamics is hence dominated by the $SU(P)$ theory which 
has $M$ flavors and generates an ADS superpotential for the meson, which is a 
field $\phi$ in the adjoint of $SU(M)$. We obtain a superpotential
\beqa
W= X_{11} \phi + (P-M)\, \left(\frac{\Lambda^{3P-M}}{\det \phi}\right)^{\frac{1}{P-M}}
\eeqa
where $\Phi$ is the original field in the adjoint of $SU(M)$. 
It is clear that there is no supersymmetric vacuum in this case, since F-term constraints cannot be satisfied.

It is also easy to realize that there is a runaway direction for large $\phi$. 
In order to make it explicit, let us restrict to the simplest case of one fractional 
brane of type $(1,0,0)$ and $P$ of type $(0,1,0)$. Then the gauge theory is just 
$SU(P)$ with one flavor, and the complete superpotential is
\beqa
W = \phi X_{11} + (P-1) \left( \frac{\Lambda^{3P-1}}{\phi} \right)^{\frac{1}{P-1}}
\eeqa
Notice that in this case the determinant in the ADS part is very simple, since $\phi$ 
is just a complex number. The F-terms are
\beqa
\frac{\partial W}{\partial X_{11}}=\phi \quad, \quad
\frac{\partial W}{\partial \phi}= X_{11} +  
\Lambda^{\frac{3P-1}{P-1}} \phi^{-\frac{P}{P-1}}
\eeqa
Clearly there is no supersymmetric vacuum. Looking for minima of the F-term scalar 
potential
\beqa
V = \phi \phi^* + (X_{11} -  \Lambda^{\frac{3P-1}{P-1}} \phi^{-\frac{P}{P-1}}) 
( X_{11}^* + (\Lambda^*)^{\frac{3P-1}{P-1}} (\phi^*)^{-\frac{P}{P-1}} )
\eeqa
and upon extremization obtain
\beqa
\frac{\partial V}{\partial X_{11}} &=& 0 \to 
 X_{11}^* + (\Lambda^*)^{\frac{3P-1}{P-1}} (\phi^*)^{-\frac{P}{P-1}}=0 \nonumber \\
\frac{\partial V}{\partial \phi} &=& 0 \to \phi^* +  \Lambda^{\frac{3P-1}{P-1}}(\frac{P}{P-1}) 
\phi^{-\frac{2P-1}{P-1}} 
( X_{11}^* + (\Lambda^*)^{\frac{3P-1}{P-1}} (\phi^*)^{-\frac{P}{P-1}} ) =0
\eeqa
Substituting the first equation into the second one we get $\phi^*=0$, and then the first 
gives $X_{11}\to\infty$. This means that there is a runaway to a minimum at infinity 
in $X_{11}$, namely the $N=2$ fractional brane runs to infinity. This agrees with the 
above physical interpretation that the fractional brane $(1,0,0)$ is thus pushed to 
infinity along this curve of singularities.

This is the first example of a situation where fractional branes which by themselves 
lead to $N=1$ supersymmetric RG flows, do not have a supersymmetric vacuum when combined. Further examples will appear in subsequent sections.

\section{Higher del Pezzo examples}
\label{higherdpn}

In this section we consider other examples of fractional branes, which in principle 
trigger cascades which are dual to the throats in \cite{Franco:2004jz}, but which do 
not fulfill the criterion in \cite{Franco:2005fd} to lead to a complex deformed 
geometry. In all cases the corresponding infrared dynamics leads to DSB.

\subsection{The $dP_2$ case}

Let us consider the cone over $dP_2$. The quiver field theory has been constructed in 
\cite{Feng:2000mi}. Diverse aspects of this theory have been studied in
\cite{Feng:2001xr,Feng:2001bn,Feng:2002zw,Feng:2002fv,Franco:2002ae,Bertolini:2004xf,
Franco:2005fd} 

\begin{figure}[ht]
  \epsfxsize = 4cm
  \centerline{\epsfbox{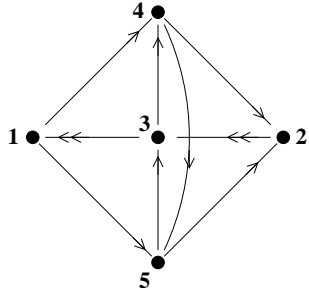}}
  \caption{\small Quiver diagram for the $dP_2$ theory.}
  \label{quiverdp2}
\end{figure}

\begin{figure}[ht]
  \epsfxsize = 8cm
  \centerline{\epsfbox{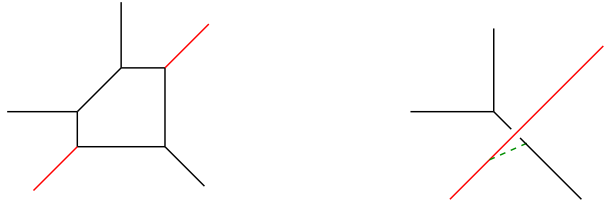}}
  \caption{\small Web diagram for the cone over $dP_2$ and its deformation.}
  \label{webdp2}
\end{figure}

The quiver for the theory is shown in figure \ref{quiverdp2}, and the superpotential 
is
\beqa
W & = & X_{34}X_{45}X_{53}-(X_{53}Y_{31}X_{15}+X_{34}X_{42}Y_{23})
\nonumber\\
  & + & 
(Y_{23}X_{31}X_{15}X_{52}+X_{42}X_{23}Y_{31}X_{14})-X_{23}X_{31}X_{14}X_{45}X_{52}
\label{W_dP2_1}
\eeqa
The web diagram is shown in figure \ref{webdp2}a.

A basis of fractional branes is $(1,1,0,0,0)$ and $(1,0,1,0,2)$. The
first fractional brane triggers a cascade which ends in a complex
deformation, as discussed in \cite{Franco:2005fd} and shown in \fref{webdp2}b. 
The second one is also expected to trigger a duality cascade, although it has not 
been studied in the literature. In any event, we now focus on the field theory 
dynamics on a set of such fractional branes. Using the criteria in 
\cite{Franco:2005fd}, the infrared dynamics does not correspond to a complex 
deformation. 

\begin{figure}[ht]
  \epsfxsize = 7cm
  \centerline{\epsfbox{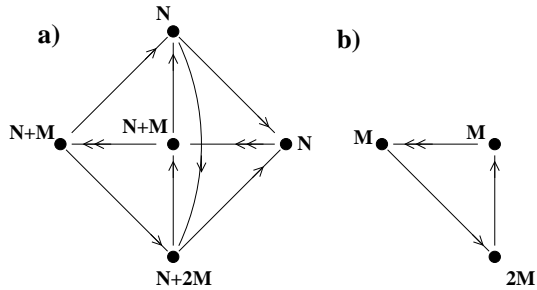}}
  \caption{\small Quiver gauge theory for the cone over $dP_2$ with $M$ fractional 
branes of the $(1,0,1,0,2)$ kind, with $N$ additional D3-branes (figure a) or by 
themselves (figure b).}
  \label{quiverdp2end}
\end{figure}

The IR theory on $M$ such fractional branes is given by the theory with ranks 
$M(1,0,1,0,2)$, shown in \fref{quiverdp2end}b. As in the previous example, it is 
easy to show that the theory does not have classical flat directions, in agreement 
with the interpretation as a fractional brane. 

Again, it is straightforward to realize that the infrared theory has DSB. We 
center on the regime where the highest rank factor dominates the dynamics. The 
$SU(2M)$ theory has $M$ flavors and develops an ADS superpotential. Hence the 
superpotential grows for large and small vevs and the theory breaks supersymmetry, 
with a non-supersymmetric minimum at finite distance (strictly speaking, only for an 
artificially fixed choice of FI terms, see section \ref{dp1}).

Similarly, when probed with a small number $N\ll M$ of additional regular D3-branes as in 
\fref{quiverdp2end}, the latter are repelled from the tip of the 
throat.

\medskip

Since the theory admits two different fractional branes, we can consider introducing a 
general linear combination of both, and consider the theory with rank vector
$(N+M+P,N+M,N+P,N,N+2P)$. The analysis in \cite{Franco:2005fd} suggests that 
confinement/complex deformation occurs only for $P=0$, and arbitrary $M$. The above 
analysis has shown DSB for $M=0$, and arbitrary $P$. In what follows we consider other 
possibilities.

One can easily argue on general grounds that a set of fractional branes involving 
both kinds leads to DSB. Namely, consider the rank vector for general fractional 
D-branes in the absence of D3-branes, $(M+P,M,P,0,2P)$, with $M,P>0$. The $SU(M+P)$ 
gauge factor has $2P$ flavors, hence there is an ADS superpotential (and 
corresponding DSB following the by now standard reasoning) for $M<P$. On the other hand, the $SU(2P)$ factor has $M+P$ flavors, 
hence there is an ADS superpotential triggering DSB for $P<M$. Hence the generic
infrared dynamics corresponds to DSB, with cases of supersymmetric complex 
deformation given by the criterion in \cite{Franco:2005fd}.

\begin{figure}[ht]
  \epsfxsize = 10cm
  \centerline{\epsfbox{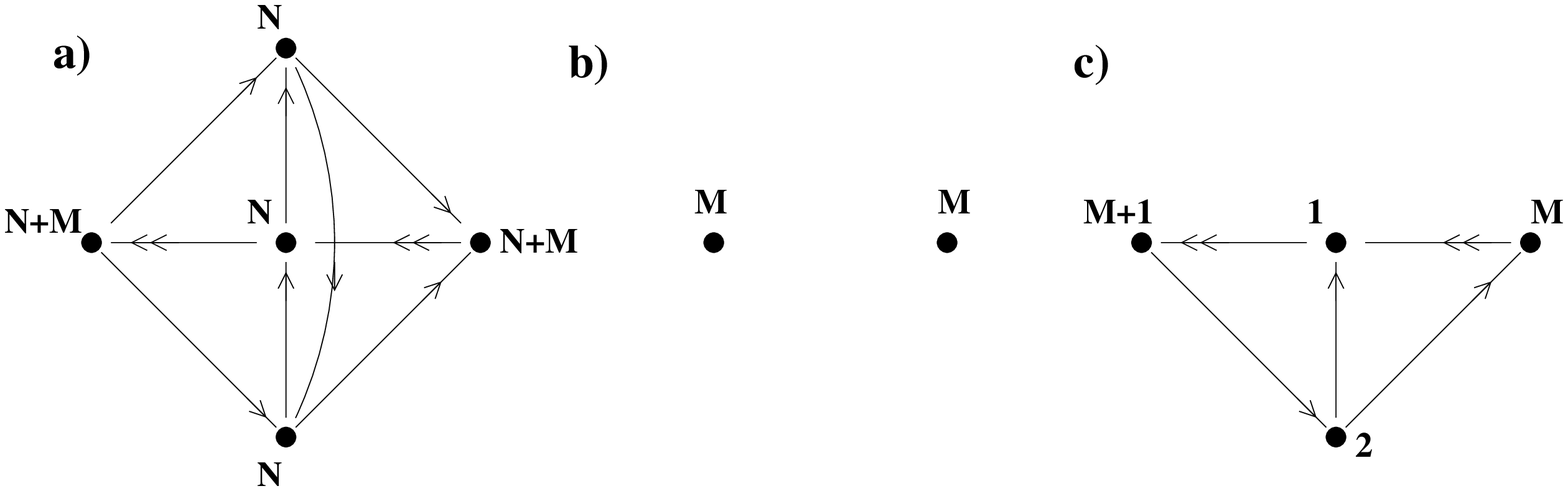}}
  \caption{\small Figures a) shows the quiver theory describing $dP_2$ with $N$ 
D3-branes and $M$ fractional branes of the $(1,1,0,0,0)$ kind, while figure b) shows 
the quiver for the infrared limit of the corresponding duality cascade. Figure c) shows 
the addition of a fractional D-brane probe of the $(1,0,1,0,2)$ kind to that 
theory.}
  \label{addingprobedp2}
\end{figure}

Let us consider an interesting particular limit.
The simplest situation is when one of the fractional brane numbers is much larger 
than the other. For instance, we can consider large $M$ and $P=1$. In this situation, 
we expect the additional fractional D-brane of the second kind to behave as a probe 
in the complex deformed throat created by the fractional branes of the first kind.
The infrared theory is shown in \fref{addingprobedp2}.c. Notice that as in previous 
sections the $U(1)$ gauge group does not appear at low energies, and simply provides 
fundamental flavors from its bi-fundamentals. Again one can argue as above to show 
that this theory breaks supersymmetry.

Notice that the qualitative picture of this system in the gravity dual is quite 
different from the previous theories. Namely we have a small number of fractional 
D-brane probes in a nice and smooth complex deformed geometry. It would be 
interesting to find out more about the nature of the gravity 
description of this dual D-brane, and the way in which supersymmetry is broken in the gravity 
picture.

\subsection{The $dP_3$ case}

Consider now the $dP_3$ theory. This theory has been constructed in  
\cite{Feng:2000mi}. Diverse aspects of this theory have been studied in
\cite{Feng:2001xr,Feng:2001bn,Feng:2002zw,Feng:2002fv,Franco:2002ae,
Franco:2005fd} 

\begin{figure}[ht]
  \epsfxsize = 4cm
  \centerline{\epsfbox{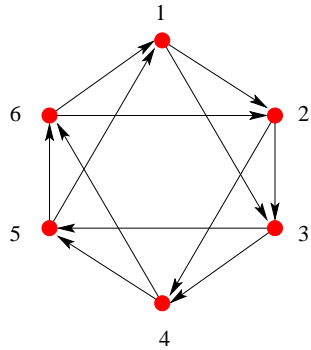}}
  \caption{\small Quiver diagram for the $dP_3$ theory}
  \label{quiverdp3}
\end{figure}

We focus on the $dP_3$ phase whose quiver is shown in \fref{quiverdp3}, with superpotential 
\beqa
W & = & X_{12} X_{23} X_{34} X_{45} X_{56} X_{61} \, +\,  X_{13} X_{35} X_{51} + 
X_{24} X_{46}
X_{62} \, - \nonumber  \\
&-& X_{23} X_{35} X_{56} X_{62}\, -\, X_{13} X_{34} X_{46} X_{61}
\, -\,  X_{12} X_{24} X_{45} X_{51}
\eeqa

\begin{figure}[ht]
  \epsfxsize = 12cm
  \centerline{\epsfbox{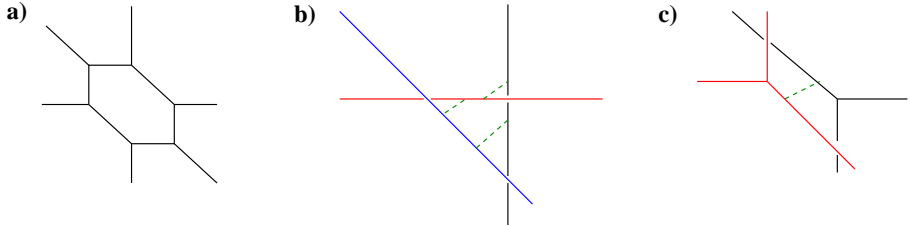}}
  \caption{\small Web diagram for the cone over $dP_3$, and its two complex 
deformations}
  \label{dp3}
\end{figure}

The web diagram is shown in figure \ref{dp3}a.

The theory admits three independent fractional branes. A basis for them   
is given by the rank vectors $(1,0,0,1,0,0)$, $(0,1,0,0,1,0)$ and $(1,0,1,0,1,0)$. 
For the general rank assignment, there is a supergravity
throat solution constructed in \cite{Franco:2004jz}, suggesting a UV cascade of
the corresponding gauge theory. The field theory analysis of 
these candidate cascades has not been carried out in the general case, but some 
examples were explicitly constructed in \cite{Franco:2005fd}.

We are interested in determining the infrared behavior of the theory for the 
general rank vector $(N+P+K,N+M,N+K,N+P,N+M+K,N)$. In \cite{Franco:2005fd} it was 
shown that for $K=0$ and arbitrary $M,P$ the cascade ends in a supersymmetric 
confining theory, whose gravity dual is described in terms of a two-parameter complex 
deformation of the geometry, as shown in figure \ref{dp3}b. On the other hand, for 
$M=P=0$ and arbitrary $K$ the cascade ends in a confining theory corresponding to the 
complex deformation in figure \ref{dp3}c.

There remains the question of the infrared behavior in the general case of 
non-vanishing $M$, $P$, $K$. Notice that $dP_3$ shows a new feature, 
compared with lower del Pezzo examples. Namely, it has the same number of 
complex deformation parameters and of independent fractional branes. One 
may be tempted to propose that the generic fractional brane leads to a 
general combination of complex deformations. However, this is clearly not 
possible, since the two complex deformations in Figures \ref{dp3}b and 
\ref{dp3}c are incompatible, hence we do not expect the infrared behavior of the 
generic fractional brane to be described by a complex deformation. 
Correspondingly, the infrared field theory does not reduce to a set of 
decoupled SYM theories without matter (or any other quiver corresponding 
to a deformation brane).

We are thus again led to discussing the dynamics of the infrared theories on 
fractional D-branes. The situation is again strongly suggestive of DSB in the 
general case. Moreover, this can be easily checked explicitly in some simple 
situations by setting to 1 the number of fractional branes of a given kind.

\begin{figure}[ht]
  \epsfxsize = 12cm
  \centerline{\epsfbox{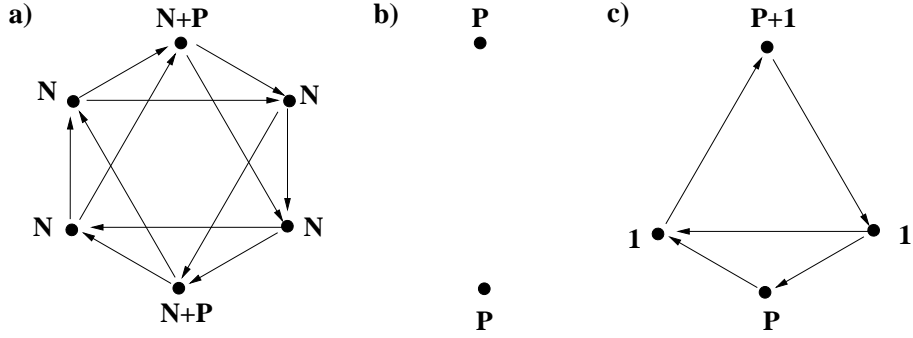}}
  \caption{\small Quiver diagrams for the theory with $P$ fractional 
branes of the $(1,0,0,1,0,0)$ kind, at a generic point in the cascade (figure a), and 
in the infrared limit (figureb). Figure c shows the infrared limit of the same theory
with an additional fractional brane of the $(1,0,1,0,1,0)$ kind.}
  \label{addingprobedp3}
\end{figure}

For instance, the regime of large $P$ and small $K$ is illustrated by taking $K=1$.
We expect this theory to have a UV cascade triggered by the $P$ fractional branes of 
the first kind (dual to a throat based on the complex deformation of the cone over 
$dP_3$ to the conifold \cite{Franco:2005fd }), with an IR influenced by the additional single brane probe.
The gauge theory is shown in figure \ref{addingprobedp3}c.
Again, it is easy to argue that this theory breaks supersymmetry dynamically. Namely, 
because we only have fractional branes, there are no classical flat directions, and all 
gauge invariants are frozen to zero. On the other hand, the $SU(P+1)$ and the $SU(P)$ 
gauge groups generate ADS superpotentials which push their mesons to non-zero
vevs.

The interpretation of the DSB on the gravity side should be similar to that for the 
$dP_2$ theory in the previous section. Namely we have a single supersymmetry 
breaking D3-brane probe in a smooth complex deformed space.

\medskip

\begin{figure}[ht]
  \epsfxsize = 10cm
  \centerline{\epsfbox{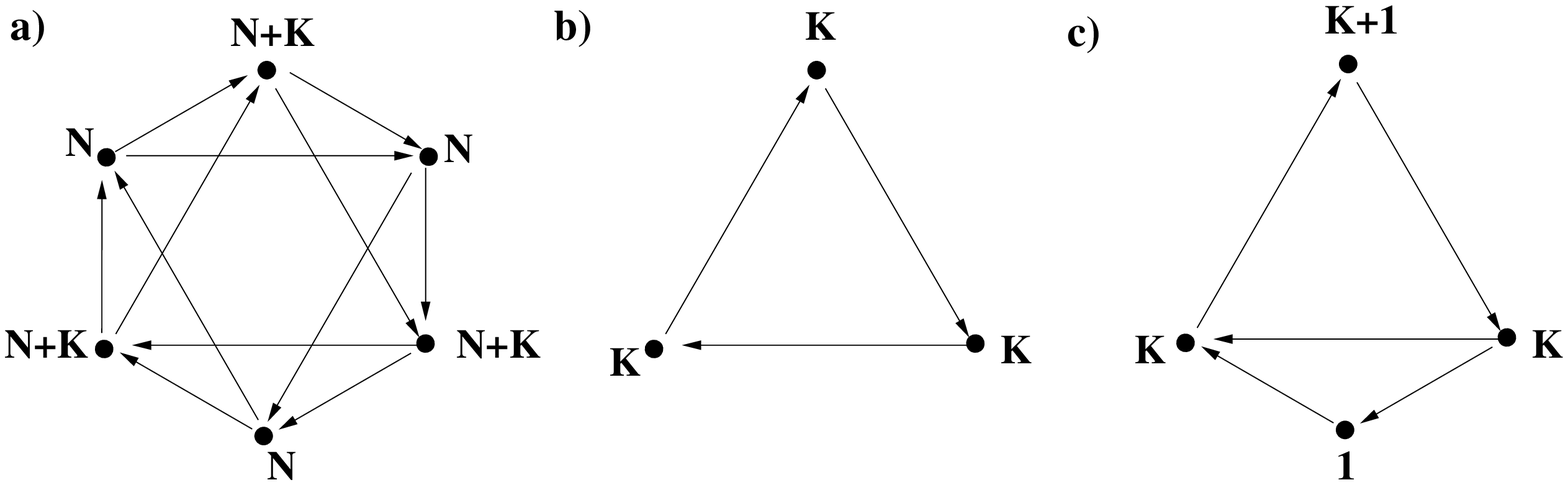}}
  \caption{\small Quiver diagrams for the theory with $K$ fractional 
branes of the $(1,0,1,0,1,0)$ kind, at a generic point in the cascade (figure a), 
and in the infrared (figureb). Figure c shows the infrared limit of the same theory
with one additional fractional brane of the $(1,0,0,1,0,0)$ kind.}
  \label{moreprobedp3}
\end{figure}

Similar conclusions hold for example in the limit of small $P$ and large $K$. Consider 
starting with large $K$ and small $M$, $P$, e.g. $M=0$, $P=1$. The infrared limit of the 
theory in the absence of the latter is shown in figure \ref{moreprobedp3}b.
Confinement of this theory triggers a complex deformation to smooth
geometry. Adding now the $P=1$ fractional brane the infrared theory is shown in 
figure \ref{moreprobedp3}c. One can easily show using by now familiar arguments 
that the theory breaks supersymmetry.

\medskip

The general conclusion is that the generic choice of fractional branes leads to DSB, 
except for the rank choices studied in \cite{Franco:2005fd}, namely those leading to 
`compatible' complex deformations.

\subsection{Beyond $dP_3$}

The above analysis can be extended to other examples, and in particular to the 
cones over toric blowups of $dP_3$ (called pseudo del Pezzos ($PdP$) in 
\cite{Feng:2002fv}). The techniques and the physical pictures are identical to those 
in the above examples, so we simply discuss a few interesting situations.

\begin{figure}[ht]
  \epsfxsize = 6cm
  \centerline{\epsfbox{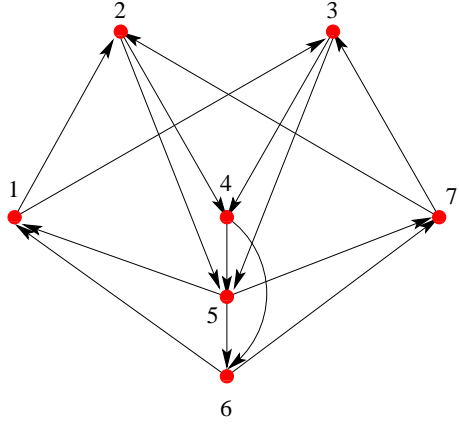}}
  \caption{\small The quiver for the $PdP_4$ theory.}
  \label{quiver_PdP4}
\end{figure}

\begin{figure}[ht]
  \epsfxsize = 4cm
  \centerline{\epsfbox{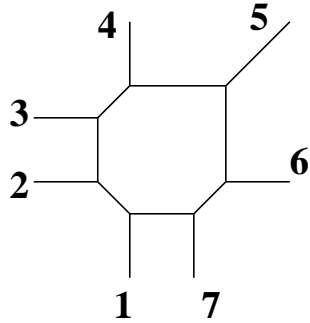}}
  \caption{\small The web diagram for the $PdP_4$ theory.}
  \label{web_PdP4I}
\end{figure}

Let us center for example on phase I of the $PdP_4$ theory in \cite{Feng:2002fv}. 
The quiver diagram is shown in figure \ref{quiver_PdP4}. The superpotential 
reads
\beq
\begin{array}{rl}
W & =X_{24}X_{46}X_{61}X_{12}+X_{73}X_{35}X_{57}-X_{73}X_{34}X_{46}X_{67}
-X_{45}X_{57}X_{72}X_{24} \\
  & 
 -X_{35}X_{56}X_{61}X_{13}+
X_{51}X_{13}X_{34}X_{45}-X_{25}X_{51}X_{12}+X_{25}X_{56}X_{67}X_{72}
\end{array}
\eeq
The web diagram is shown in figure \ref{web_PdP4I}

A basis of fractional branes in this theory is provided by the rank vectors 
$(1,0,0,1,0,0,0)$, $(0,1,0,0,0,1,0)$, $(1,1,0,0,1,0,0)$ and $(1,0,1,0,1,0,0)$. The 
dynamics for the corresponding field theories are easy to identify following the 
discussion in Section \ref{section_classification_fractional}. For instance, the 
fractional branes $(1,0,0,1,0,0,0)$ and $(0,1,0,0,0,1,0)$ lead to quivers with 
decoupled nodes, hence lead to confinement triggering a complex deformation in the 
dual geometry (described by recombination of the legs 1 and 4, resp. 2 and 6, in the 
web diagram). The fractional brane $(1,1,0,0,1,0,0)$ leads to a quiver with three 
nodes joined by a loop of arrows, with the gauge invariant $X_{25}X_{51}X_{12}$ 
appearing in the superpotential. Hence it also corresponds to a deformation brane, producing the complex deformation obtained by recombining the legs 1, 2 
and 5 in the web diagram. Finally, the fractional brane $(1,0,1,0,1,0,0)$ leads to a 
quiver with three nodes joined by a loop of arrows, but with the gauge invariant 
$X_{35}X_{51}X_{13}$ {\em not} appearing in the superpotential. Hence it is an $N=2$ 
fractional brane. Notice that this is despite the fact that the web diagram suggests 
a complex deformation by recombination of the legs 1, 2 and 5. This is due to the 
familiar fact that web diagrams with parallel external legs sometimes miss important 
information concerning the superpotential of the quiver theories \cite{Feng:2004uq}, and it is 
precisely the presence or not of certain superpotential terms that distinguishes 
certain $N=2$ branes from deformation branes.

It is easy to construct other examples of fractional branes with nice supersymmetric 
infrared dynamics. Since cones over these higher del Pezzos contain many 2-cycles, 
equivalently fractional branes, the pattern is very rich. Indeed, there are 
new situations not present in previous examples. For instance, the theory given 
by the rank vector $(M,P,M,0,M,P,0)$ describes $M$ $N=2$ branes of type 
$(1,0,1,0,1,0,0)$ and $P$ deformation branes of type $(0,1,0,0,0,1,0)$. The two are 
compatible, in the sense that the complex deformation does not remove the curve of 
$A_1$ singularities on whose collapsed 2-cycle the $N=2$ fractional branes are 
wrapped. Hence the field theory should be supersymmetric and have a one-dimensional 
flat direction. This is indeed the case, upon careful analysis of the quiver field 
theory. 

\begin{figure}[ht]
  \epsfxsize = 3cm
  \centerline{\epsfbox{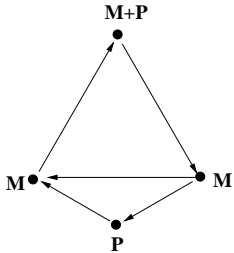}}
  \caption{\small Quiver for certain fractional brane in the $PdP_4$ theory.}
  \label{quiverpdp4}
\end{figure}

It is also easy, and in fact very generic, to have rank vectors which lead to DSB. 
Again, the pattern is very intricate, and we simply mention a few examples, which 
illustrate a particular point not manifest in previous theories. This is the key 
role played by the superpotential in the pattern of DSB, as illustrated in the 
following by two fractional brane theories with identical quiver but different 
superpotentials. Consider for instance the rank vector $(M+P,M,0,P,M,0,0)$. The 
quiver diagram is shown in figure \ref{quiverpdp4}, and the superpotential is
\beqa
W  = -X_{25}X_{51}X_{12}
\eeqa
The theory does not have classical flat directions, since the F-term equation 
$X_{51}X_{12}=0$ lifts the D-flat directions parametrized by vevs for the gauge 
invariants $X_{25}X_{51}X_{12}$ and $X_{51}X_{12}X_{14}X_{45}$. At the quantum 
level, the $SU(M+P)$ gauge factor develops an ADS superpotential, which thus leads 
to a supersymmetry breaking minimum, using by now familiar arguments.

It is interesting to compare the situation with the theory given by the rank vector 
$(M+P,0,M,P,M,0,0)$. The quiver diagram also corresponds to figure \ref{quiverpdp4}, 
but with node labels different from above. The superpotential is also different and 
reads
\beqa
W  = X_{51}X_{13}X_{34}X_{45}
\eeqa
Now there is a classical flat direction parametrized by $X_{35}X_{51}X_{13}$. At the 
quantum level, the $SU(M+P)$ gauge factor develops an ADS superpotential, which 
lifts the flat direction turning it into a runaway direction.

Clearly it is easy to construct many other examples with DSB patterns similar to 
those that appeared in previous theories. We leave a more systematic discussion as an 
exercise for the interested reader.

\section{The $Y^{p,q}$ family}
\label{ypqfamily}

Recently an infinite class of Sasaki-Einstein 5d metrics on $S^2\times S^3$ has been 
constructed, denoted $Y^{p,q}$, which can be used to build an infinite class of 6d 
conical Calabi-Yau geometries  
\cite{Gauntlett:2004zh,Gauntlett:2004yd,Gauntlett:2004hh,Gauntlett:2004hs,Martelli:2004wu}.
The dual quiver field theories on D3-brane on such singularities have been 
proposed in \cite{Benvenuti:2004dy,Benvenuti:2004wx}. The impressive matching between 
the field theory results and the the geometry in this family is exemplified by the 
matching of R-charges and volumes, which interestingly are in general irrational as 
follows by $a$-maximization \cite{Bertolini:2004xf,Benvenuti:2004dy,Martelli:2005tp}.

The $Y^{p,q}$ theories thus provide an infinite class of examples to test new ideas 
in quiver gauge field theories with explicit gravity duals. Moreover, they can be 
exploited as in \cite{Hanany:2005hq}, to generate new infinite pairs of quivers/geometries by un-Higgssing.

In \cite{Ejaz:2004tr}, warped throat solutions with 3-form fluxes were explicitly 
constructed, based on the conical metrics of $Y^{p,q}$, and hence containing a naked 
singularity at their origin. These throats are expected to be dual to the RG flow 
of the dual field theory in the presence of fractional branes. In fact in 
\cite{Ejaz:2004tr} explicit duality cascades were described for particular choices 
of $p,q$. In this section we discuss the infrared behavior of the $Y^{p,q}$ 
theories in the presence of fractional branes.

\begin{figure}[ht]  
  \epsfxsize = 10cm   
  \centerline{\epsfbox{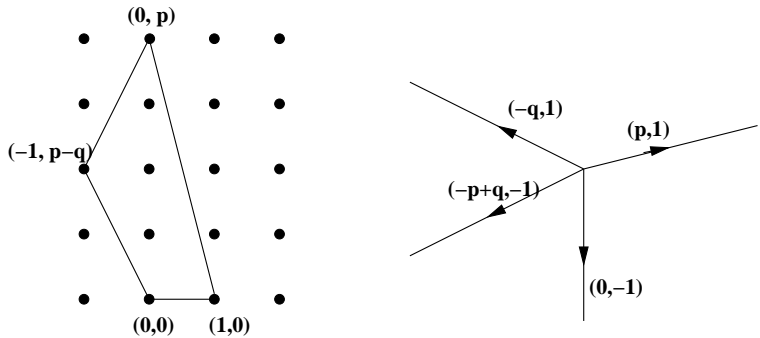}}
  \caption{The toric and web diagram for the cone over the general
$Y^{p,q}$ manifold. No leg recombination is possible except for the case
$q=0$.}
  \label{ypq} 
\end{figure}

As discussed in \cite{Franco:2005fd}, the $Y^{p,q}$ theories do not admit complex 
deformations, except for $Y^{p,0}$ (which is simply a $\IZ_p$ quotient of the 
conifold). This follows from the fact that the corresponding web diagram (shown in 
figure \ref{ypq}) does not admit a splitting into subwebs in equilibrium.
As in section \ref{dp1}, this would seem to contradict the recent explicit 
construction of a first order complex deformation in \cite{Burrington:2005zd}. 
However it is possible to use mathematical results on complex deformations for 
toric singularities (see appendix \ref{obstruction}) to show that a first order 
deformation indeed exists, but is obstructed at second order. 

On the other hand, $Y^{p,p}$ is an orbifold of 
flat space, and the fractional brane corresponds to an $N=2$ subsector, thus its 
infrared dynamics is dominated by enhan\c{c}on behaviors 
\cite{Johnson:1999qt,Bertolini:2000dk,Gauntlett:2001ps}.

\begin{figure}[ht]
  \epsfxsize = 8cm
  \centerline{\epsfbox{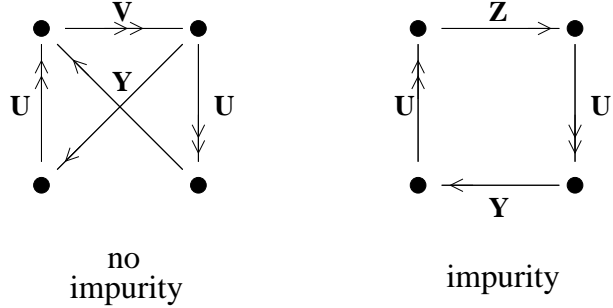}}
  \caption{\small No-impurity and impurity fundamental cells for $Y^{p,q}$ quivers that only contain single impurities.}
  \label{sigma_tau_cells}
\end{figure}

Following our general proposal, we claim that there is DSB for any of the $Y^{p,q}$ 
theories $0<q<p$. To show this we simply need to consider the infrared theory where 
only fractional branes are present. In addition, we may choose any toric phase of the 
quiver field theory, since all are related by Seiberg duality, and hence have 
equivalent infrared dynamics. The toric \footnote{Here we use the notation 
introduced in \cite{Feng:2002zw}, where a quiver is denoted 'toric' when all 
the gauge groups have the same rank when only probe D3-branes are included.}
Seiberg dual phases of $Y^{p,q}$ quivers were fully classified in 
\cite{Benvenuti:2004wx}. There, it was shown that they can be constructed by 
modifying $\IC^3/\IZ_{2p}$ quivers with so called single and double impurities. The 
effect of Seiberg duality is to move impurities around the quiver and double 
impurities are produced whenever single impurities collide. Out of the toric phases in 
\cite{Benvenuti:2004wx}, we focus on those with only single impurities. We can construct 
them by combining two building blocks, which we call no-impurity cell and impurity cell \footnote{This construction was originally cenceived by Pavlos Kazakopoulos.}. 
We show them in \fref{sigma_tau_cells}.

\begin{figure}[ht]
  \epsfxsize = 12cm
  \centerline{\epsfbox{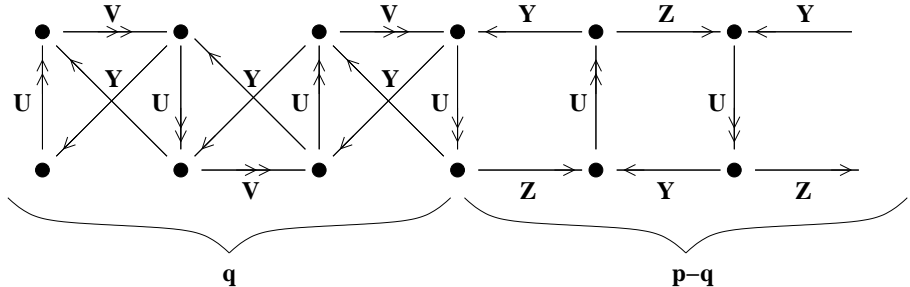}}
  \caption{\small General picture of the toric phase of $Y^{p,q}$ that we
are considering}
  \label{figypq}
\end{figure}

We find it convenient to choose the toric phase with $p-q$ 
single impurities next to each other (namely, a sequence of $q$ no-impurity cells 
followed by $p-q$ impurity cells). This is shown in figure \ref{figypq}, 
where the nodes are periodically identified after $p$ cells.

We would like to consider the theory in the presence of fractional branes.
In order to give the result of the corresponding rank vector, we choose to order the 
nodes starting with that in the lower left, and continuing the sequence of nodes 
by following the bifundamentals $U,V,U,V\ldots$ ($q$ times), and then 
$U,Z,U,Z\ldots$ ($p-q$ times). Following the discussion in 
\cite{Benvenuti:2004wx}, the rank vector for fractional branes can be obtained by 
using the baryonic charges of the diverse bifundamental fields among the nodes. 
The charges for the bifundamentals in $Y^{p,q}$ theories are as follows
\cite{Benvenuti:2004dy}: $U$ has charge $-p$, $V$ has charge $q$, $Z$ has charge 
$p+q$ and $Y$ has charge $p-q$. Using the ordering of nodes described above, the 
rank vector is
\beq
\begin{array}{ll}
\vec{N} & = p (0,1,1,2,2,3,3, \ldots, q, q, (q+1);q,(q+1),q,(q+1),\ldots,q,(q+1)) \\
  & - q(\underbrace{0,0,1,1,2,2, \ldots, (q-1), q, 
q}_{2(q+1)};\underbrace{(q+1),(q+1),(q+2),(q+2),\ldots,(p-1),(p-1)}_{2(p-q-1)})
\end{array}
\label{general_ranks}
\eeq

In a similar spirit, we can take the general version of the quiver, and compute a 
general expression for a vector $\vec{N_F}$, giving the numbers of flavors for each 
node,

{\footnotesize
\beq
\vec{N}_F=(\underbrace{N_{2p}+N_3, 2N_1+N_4, 2N_2+N_5,\ldots}_{2q};2N_{2q}+N_{2q+4},2N_{2q+1};
\underbrace{2N_{2q+4},2N_{2q+3},2N_{2q+6},2N_{2q+5},\ldots,2N_{2p-1}}_{2(p-q-1)})
\eeq
}
Notice that we have separated the $(2q+1)$- and $(2q+2)$-th nodes, which lie near 
the boundary of the impurity region in the quiver. The first node is also special. 
Using the general ranks in \eref{general_ranks}, we get

{\footnotesize
\beq
\begin{array}{ll}
\vec{N}_F & = p (2,2,4,5,7,8,10,11 \ldots, (3q-2), 
(3q-1);(3q+1),2q;2(q+1),2q,2(q+1),2q,\ldots,2q) \\
  & - q(\underbrace{0,1,2,4,5,7,8,\ldots, 
(3q-4),(3q-2)}_{2q};(3q-1),2q;\underbrace{2(q+1),2(q+1),2(q+2),2(q+2),\ldots,2(p-1)}_{2(p-q-1)})
\end{array}
\label{general_flavors}
\eeq
}
Using \eref{general_ranks} and \eref{general_flavors}, we get

{\footnotesize
\beq
\begin{array}{ll}
\vec{N}_F-\vec{N}= & = p (2,1,3,3,5,5,\ldots, (2q-1), 
(2q-1);(2q+1),(q-1);(q+2),(q-1),(q+2),(q-1),\ldots,(q-1)) \\
  & - q(\underbrace{0,1,1,3,3,5,5,7, \ldots, 
(2q-3),(2q-1)}_{2q};(2q-1),q;\underbrace{(q+1),(q+1),(q+2),(q+2),\ldots,(p-1)}_{2(p-q-1)}) 
 \end{array} \nonumber
\label{general_difference}
\eeq
}

We can now analyze this result. Consider the case $0< q < p$. We realize that 
the last node has 
\beqa
N_F-N_c\, =\, p(q-1)-q(p-1)\, =\, -p+q
\eeqa
and hence generates an ADS superpotential. Combining it with the classical superpotential in \cite{Benvenuti:2004dy}, we conclude that fractional 
branes trigger DSB for the whole family of $Y^{p,q}$ theories (for $0<q<p$). 
This is in agreement with the remark of absence of complex deformations in 
\cite{Franco:2005fd}.

Note that for $p=q$, only the first $2p$ entries in \eref{general_difference} are 
present. All of them are greater or equal to zero and and we see that there is no 
DSB. This theory corresponds to an orbifold of flat space, and the fractional brane 
is of $N=2$ type, hence leading to supersymmetric infrared dynamics. On the other 
hand, the case $q=0$ is also special, and was analyzed in \cite{Franco:2005fd}.
The infrared behavior of these theories is complex deformation, and is related to that 
of the conifold by a $\IZ_p$ orbifold action.

\medskip

We conclude by showing how our expressions work in an explicit example, 
concretely the $Y^{4,2}$ theory. In this case, the ranks
and flavors become

\beq
\vec{N}=(0,4,2,6,4,8,2,6) \ \ \ \ \ \ \vec{N_F}=(8,6,12,12,18,8,12,4)
\eeq

Then,

\beq
\vec{N}_F-\vec{N}_c=(8,2,10,6,14,0,10,-2)
\eeq

And we conclude that node 8 develops an ADS superpotential.

It would be nice to extend this kind of general analysis to other infinite families 
of toric singularities, like the $X^{p,q}$ theories in \cite{Hanany:2005hq}. In the 
meantime, it is straightforward to carry out the analysis in many different 
concrete examples.

\section{Brane tilings and fractional branes}
\label{brane_tiling}

In \cite{Franco:2005rj} a new type of brane configurations dual to gauge theories 
on D3-branes probing arbitrary toric singularities was introduced. They were named 
{\bf brane tilings} and encode both the quiver and the superpotential of the gauge 
theory. Brane tilings provide a connection to dimer models (see \cite{Hanany:2005ve}
for an introduction in the present context) and considerably simplify the study of 
these theories. In particular, the previously laborious task of finding the moduli 
space of the gauge theory is reduced to the computation of the determinant of the 
Kasteleyn matrix of a graph.

A brane tiling consists of an NS5-brane spanning the 0123 directions and wrapping an
holomorphic curve in 4567. A simple way to visualize this configuration is by 
considering the intersection of this curve with the 46 plane, where it resembles a 
tiling. In addition to the NS5-brane, D5-branes extend in the 0123 directions and 
are finite in 46, being suspended from the NS5-brane like soap bubbles, and filling 
the faces of the tiling. The 4 and 6 coordinates are periodically identified, i.e. 
the configuration lives on the surface of a 2-torus.

The projection on the 46 plane defines a bipartite graph. There is an explicit mapping
between the tiling and the corresponding gauge theory. Faces in the graph are mapped to 
gauge groups, edges to bifundamental fields and nodes to superpotential terms. We refer 
the reader to \cite{Franco:2005rj} for a detailed exposition of the correspondence.

\subsection{Brane tiling perspective of fractional branes}

It is natural to address the issue of fractional branes in the brane tiling context. 
As we have discussed, fractional branes correspond to anomaly free rank assignments
in the quiver. The number of D5-branes in each face of the tiling gives the rank of 
the associated gauge group. Thus, rank vectors involving only 0 or 1 entries can be
represented by coloring the brane tiling in a two color `chessboard' fashion.
The classification of fractional branes outlined in Section 
\ref{section_classification_fractional} becomes very intuitive under this new light \footnote{Similar considerations for DSB in brane setups were made in \cite{Hanany:1997tb}.}. 
Using the gauge theory/tiling dictionary of \cite{Franco:2005rj}, the different 
types of fractional branes become:

\bigskip

\noindent{\bf `Deformation' fractional branes:} there are two types of deformation 
branes. They can correspond to either decoupled $SU(N)$ gauge group without flavors 
(set of isolated faces, possibly touching each other at nodes in the tiling) or to 
closed loop in the quiver with the corresponding term present in the superpotential 
(clusters of faces surrounding a given node in the tiling). These two types of 
chessboard configurations are by construction anomaly free. 

\bigskip

\noindent{\bf $N=2$ fractional branes:} in this case the colored faces in the tiling 
form `strips'. Every shaded face has an even number of shaded neighbors, located 
such that they contribute an equal number of incoming and outcoming arrows into the 
face, rendering the configuration anomaly free. These strips map to closed
loops in the quiver that are not contained in the superpotential.

\bigskip

\noindent{\bf DSB branes:} 
All the other anomaly free assignments of ranks corresponds to DSB branes. Moreover,
DSB branes involve rank vectors in which the non-zero entries are not all identical,
and thus they cannot be represented by simple two color shadings of the tiling. The 
non-abelian dynamics associated to some of these higher-rank gauge factors results 
in DSB as described in previous sections.

\medskip

\begin{figure}[ht]
  \epsfxsize = 5cm
  \centerline{\epsfbox{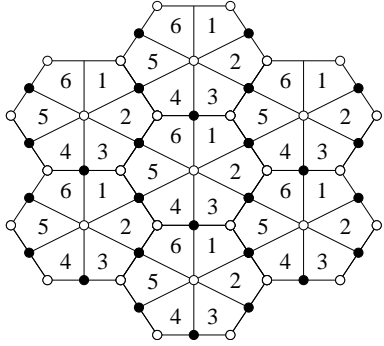}}
  \caption{\small Brane tiling for model I of $dP_3$.}
  \label{dimer_dP3_1}
\end{figure}

Let us illustrate these ideas with some explicit examples. We start with model I of $dP_3$. 
The brane tiling for this theory was introduced in \cite{Franco:2005rj}, and we present it in 
\fref{dimer_dP3_1}.

\begin{figure}[ht]
  \epsfxsize = 17cm
  \centerline{\epsfbox{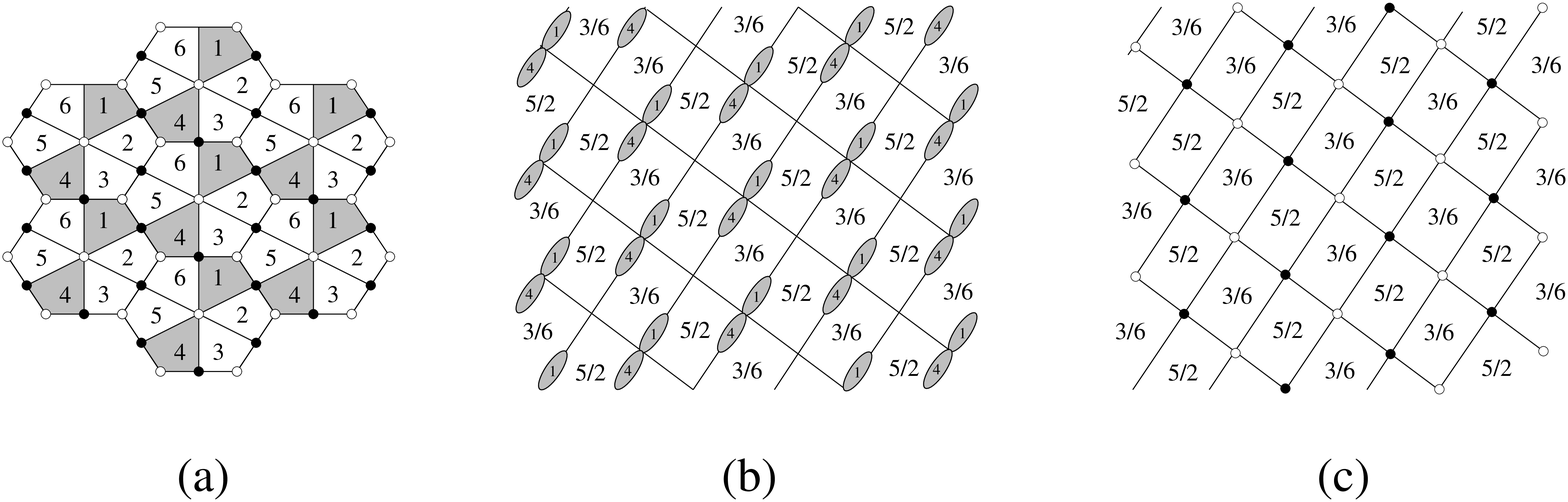}}
  \caption{\small a) Brane tiling for the $(1,0,0,1,0,0)$ deformation fractional 
brane of model I of $dP_3$. b) and c) Complex deformation to the conifold.}
  \label{dimer_fractional_dP3_1_2}
\end{figure}

This theory has a fractional brane given by the rank vector $(1,0,0,1,0,0)$ which
triggers a complex deformation in the gravity dual. Painting the faces in the  
tiling accordingly, we obtain \fref{dimer_fractional_dP3_1_2}.a. The configuration 
is given by isolated faces in the tiling and thus corresponds to the first type of 
deformation fractional branes. There is a similar fractional brane given e.g. by a 
rank vector $(0,1,0,0,1,0)$. Its associated tiling configuration is identical
up to a rotation so we do not present it.

Brane tilings also provide a useful way of visualizing IR deformations. We can think 
about the process as a recombination of faces in the tiling, reflecting the higgssing 
of the corresponding gauge groups due to meson vevs (so that faces 2 and 5 combine 
into a face 5/2, and similarly for 3 and 6), and the subsequent removal of the 
shaded tiles, due to confinement. \fref{dimer_fractional_dP3_1_2}.bc represents the 
deformation that takes place for an equal number of regular D3-branes and 
$(0,1,0,0,1,0)$ fractional branes, leading to the brane tiling of the conifold 
theory. This type of figure should not be interpreted as an exact step by step 
description of the dynamical process, but as a helpful bookkeeping diagram. 

\begin{figure}[ht]
  \epsfxsize = 17cm
  \centerline{\epsfbox{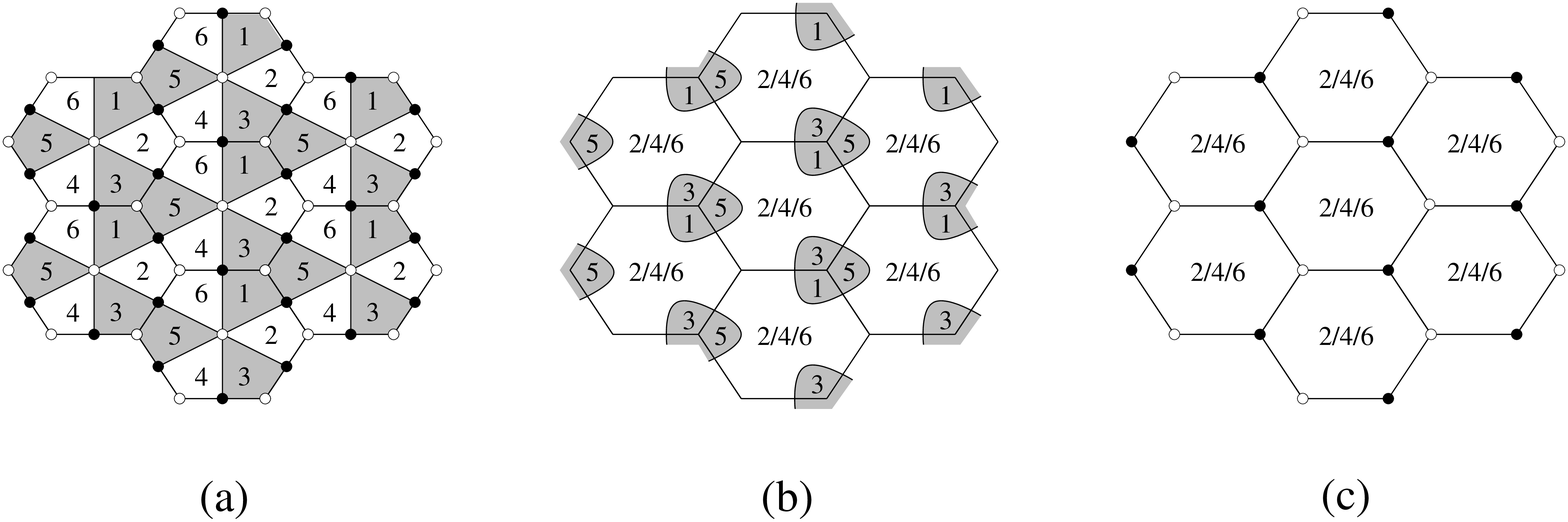}}
  \caption{\small  a) Brane tiling for the $(1,0,1,0,1,0)$ deformation fractional brane of model I of $dP_3$. b) and c) 
Complex deformation to $\IC^3$.}
  \label{dimer_fractional_dP3_1_4}
\end{figure}

There is also a another deformation brane, in this case of the second type, given 
by the rank vector $(1,0,1,0,1,0)$. \fref{dimer_fractional_dP3_1_4}.a shows the 
associated tiling. Once again there is another fractional brane, given by the vector 
$(0,1,0,1,0,1)$, which is identical to the previous one under a rotation. In this 
case, a similar picture of faces of the tiling being combined by meson vevs can be 
used to describe the deformation as shown in \fref{dimer_fractional_dP3_1_4}.bc. 
After the process, we recover the brane tiling of the flat space theory, as 
expected.

\medskip

\begin{figure}[ht]
  \epsfxsize = 12cm
  \centerline{\epsfbox{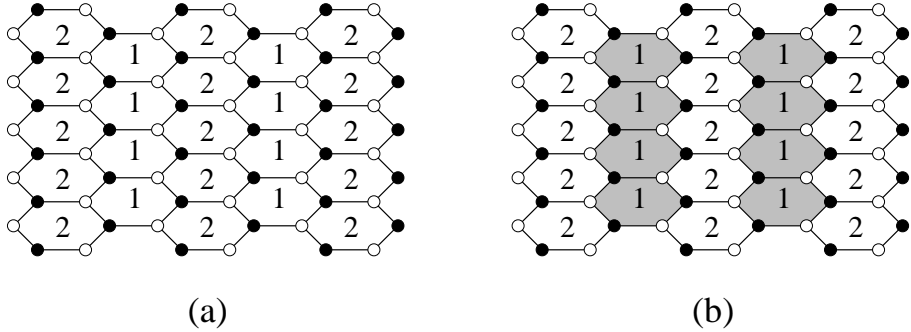}}
  \caption{\small a) Brane tiling for $\IC^2/\IZ_2$. b) Tiling representation of the $(1,0)$ $\mathcal{N}=2$ fractional brane.}
  \label{dimer_fractional_Z2}
\end{figure}

Let us now consider an example of $\mathcal{N}=2$ fractional branes. 
\fref{dimer_fractional_Z2}.a shows the brane tiling for the $\IC^2/\IZ_2$ orbifold. 
The fractional brane associated to the $(1,0)$ rank vector is obtained by painting 
the strip of all faces corresponding to node 1 as in \fref{dimer_fractional_Z2}.b. 
The modulus corresponds to recombining all painted faces by the vev of the adjoint, 
and to sliding the recombined strip off the picture, suspended between the NS branes 
on the boundary of the strip.

\begin{figure}[ht]
  \epsfxsize = 5cm
  \centerline{\epsfbox{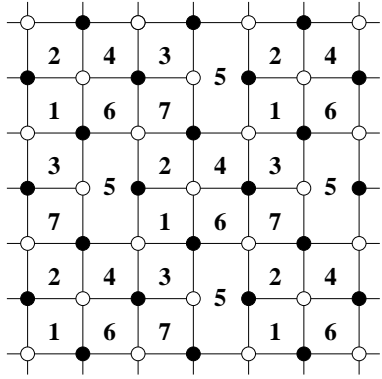}}
  \caption{\small Brane tiling for model I of $PdP_4$.}
  \label{dimer_PdP4}
\end{figure}

We now consider model I of $PdP_4$, which has both deformation branes and $\mathcal{N}=2$ branes.
The quiver diagram and superpotential for this theory can be found in \cite{Feng:2002fv}. 
Its brane tiling can be constructed according to the rules in \cite{Franco:2005rj} and is 
shown in \fref{dimer_PdP4}. 

\begin{figure}[ht]
  \epsfxsize = 16cm
  \centerline{\epsfbox{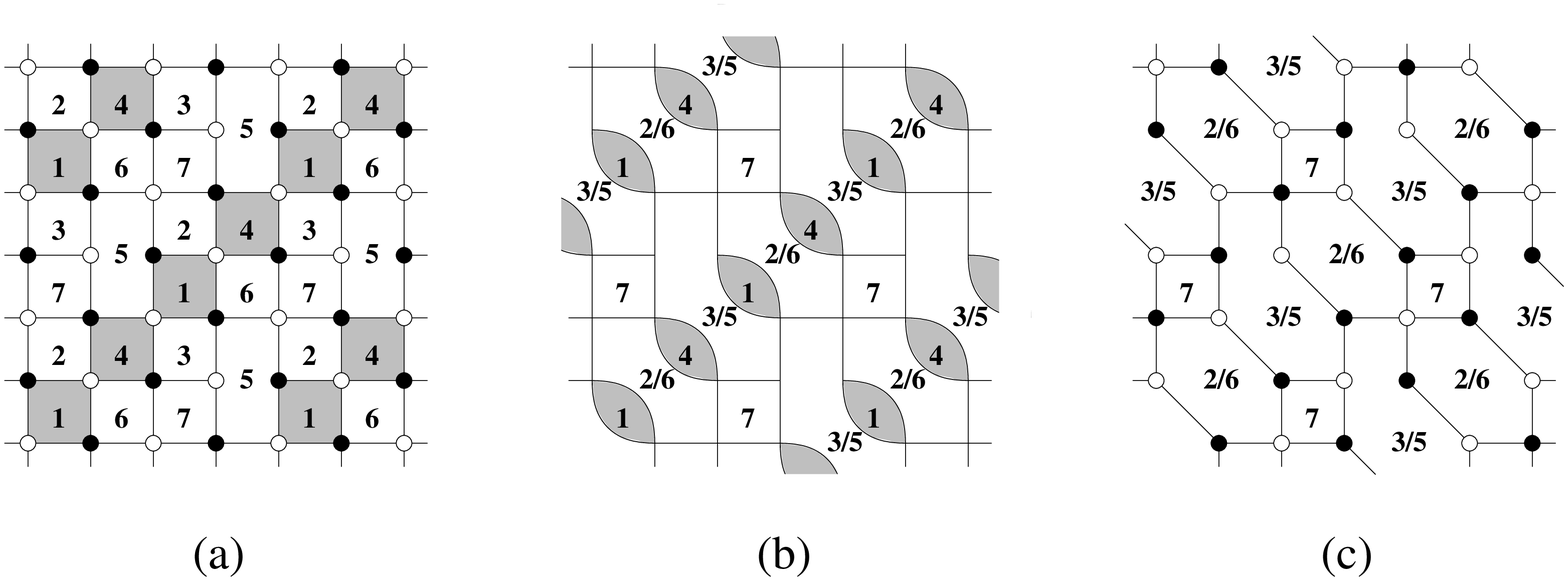}}
  \caption{\small  a) Brane tiling for the $(1,0,0,1,0,0,0)$ deformation fractional brane of model I of $PdP_4$. b) and c) Complex
deformation to the $SPP$.}
  \label{dimer_deformation_PdP41}
\end{figure}

There are different kinds of fractional branes, and we 
consider some interesting ones in the following. The $(1,0,0,1,0,0,0)$ fractional brane
is shown in \fref{dimer_deformation_PdP41}.a. It is a deformation fractional brane.
There are other similar fractional branes, related to it by symmetry, like $(0,1,0,0,0,1,0)$.

As in the above examples, we can carry out the deformation at the level of the  
brane tiling. For this fractional brane, we obtain a deformation to the $SPP$ 
tiling as shown \fref{dimer_deformation_PdP41}.bc. This agrees with the findings in 
\cite{Franco:2005fd}. Interestingly, the tiling picture even captures the 
appearance of massive fields (corresponding to 2-valent nodes \cite{Franco:2005rj}) 
which have to be integrated out using their equations of motion to arrive at the IR 
theory.

\begin{figure}[ht]
  \epsfxsize = 16cm
  \centerline{\epsfbox{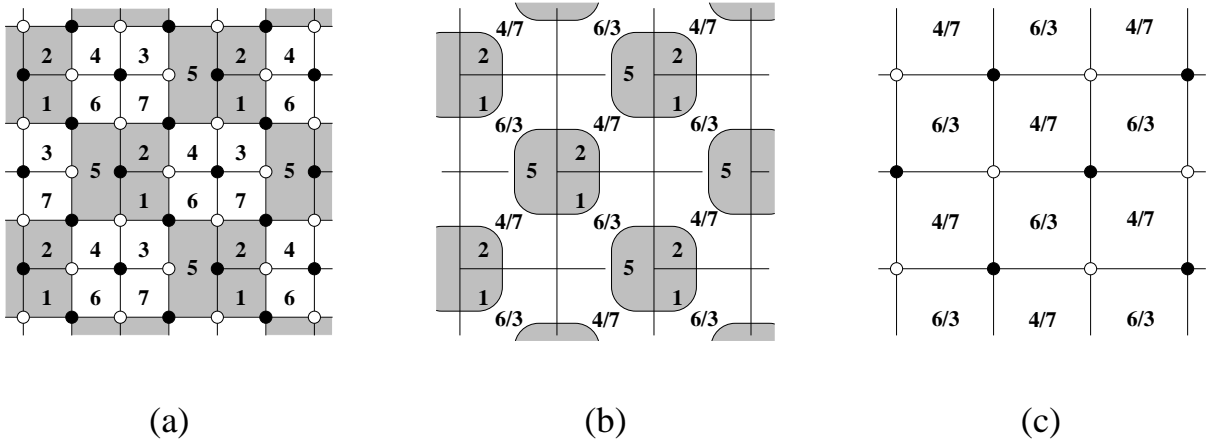}}
  \caption{\small a) Brane tiling for the $(1,1,0,0,1,0,0)$ deformation fractional brane of model I of $PdP_4$. b) and c) Complex deformation
to the conifold.}
  \label{dimer_deformation_PdP42}
\end{figure}

\fref{dimer_deformation_PdP42}.a shows the $(1,1,0,0,1,0,0)$ deformation fractional 
brane. The corresponding deformation, leading to the conifold tiling, is shown in 
\fref{dimer_deformation_PdP42}.bc. This agrees with the results of 
\cite{Franco:2005fd}. 

\begin{figure}[ht]
  \epsfxsize = 5cm
  \centerline{\epsfbox{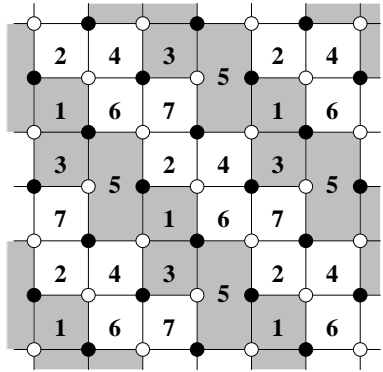}}
  \caption{\small Brane tiling for the $(1,0,1,0,1,0,0)$ $\mathcal{N}=2$ fractional brane of model I of $PdP_4$.}
  \label{dimer_fractional_PdP43}
\end{figure}

Finally, the $(1,0,1,0,1,0,0)$ brane, shown in \fref{dimer_fractional_PdP43}, is an 
$\mathcal{N}=2$ fractional brane. Similarly to what happens in the $\IC^2/\IZ_2$ 
case, its modulus parametrizes the possibility of detaching the strip of painted 
faces and sliding it along the NS5-brane along the boundary of the strip.

We conclude that brane tilings are a useful tool for the study of deformation
and $\mathcal{N}=2$ fractional branes. The left over tiling encodes
the result of the strong dynamics they trigger. A brane tiling representation
of DSB fractional branes is admittedly more complicated, since they involve more than
two different ranks. We nevertheless hope interesting progress in understanding 
the physics of DSB fractional branes using brane tiling techniques.

\subsection{Beta function relations and baryonic $U(1)$ symmetries from brane tilings}

There is a close relation between fractional branes and baryonic $U(1)$ symmetries 
in quiver gauge theories. In this section we exploit the brane tiling picture of 
fractional branes to describe a subset of these $U(1)$'s.

In a super conformal field theory, the $U(1)$ R-symmetry can in principle mix with 
every anomaly free $U(1)$ global symmetry that commutes with charge conjugation. 
In other words, in the presence of $n$ additional $U(1)$'s, the space of 
R-charges satisfying the constraints that all beta functions (i.e beta 
functions for gauge and superpotential couplings) vanish is 
$n$-dimensional. This apparent freedom in the choice of the R-charge 
is fixed by the a-maximization principle \cite{Intriligator:2003jj}.

For gauge theories constructed using brane tilings, there is one beta function for 
each face (gauge coupling $g_I$) and for each node (superpotential coupling 
$h_\alpha$). They are given, upto multiplicative factors, by

\beq
\begin{array}{lcl}
\beta_{h_\alpha} = \sum_i R_i -2 & \ \ \ \ & \mbox{i $\in$ edges ending on node $\alpha$} \\
\beta_{g_I} = 2+\sum_i(R_i-1)    & \ \ \ \ & \mbox{i $\in$ edges around face $I$}
\end{array}
\eeq

Hence, the total number of beta functions is 

\beq                                                                                
N_g+N_w=N_f
\eeq
where $N_g$ is the number of gauge groups, $N_w$ the number of superpotential 
terms and $N_f$ the number of bifundamental fields. The above equality follows
from Euler's formula applied to the tiling brane, taking into account that it lives
on the surface of a 2-torus \cite{Franco:2005rj}. The number of beta functions 
is then, a priori, equal to the number of fields. We conclude that any additional 
dimension in the space of R-charges solving the vanishing of the beta functions 
(which we saw corresponds to a $U(1)$) must come from a non-trivial relation 
between some of the beta functions.
 
There is an important subset of global $U(1)$ symmetries, denoted {\bf baryonic}.
They are gauge symmetries in the $AdS \times X_5$ gravity dual, with their corresponding 
gauge bosons coming from the reduction of the RR $C^{(4)}$ over compact 3-cycles in 
$X_5$ \cite{Intriligator:2003wr}. They are called baryonic because dibaryon operators 
in the gauge theory are dual to D3-branes wrapping 3-cycles in $X_5$ and are thus 
naturally charged under them. It is important not to confuse these baryonic symmetries 
with the baryon number, which does not commute with charge conjugation and thus 
cannot mix with the R-symmetry. There is one baryonic $U(1)$ for each fractional 
brane. Intuitively this is because fractional branes correspond to D5-branes wrapped 
over 2-cycles in $X_5$ and $b_2(X_5)=b_3(X_5)$. More technically, because one can 
use the baryonic charges to define the rank vector of a fractional brane 
\cite{Benvenuti:2004wx}.

There can be additional global $U(1)$ symmetries that can also mix with the R-charge,
arising from isometries of $X_5$, but they are not related to any fractional brane. 
The prototypical example is given by the $dP_1$ (or $Y^{2,1}$) theory 
\cite{Bertolini:2004xf} and the general $Y^{p,q}$ quivers \cite{Benvenuti:2004dy}, 
for which there is a single type of fractional brane, but there are two global 
$U(1)$'s in addition to the R-symmetry (the one usually referred to as $U(1)_F$ 
\cite{Benvenuti:2004dy} is the one coming from isometries).

Combining the statements above we see that every fractional brane implies a 
non-trivial relation among beta functions. Brane tilings give a straightforward way 
to identify the beta function combinations associated to deformation and 
$\mathcal{N}=2$ fractional branes. For each chessboard coloring of the tiling we 
have
                   
\beq                                                             
\sum_I \beta(g_I)- \sum_\alpha c_\alpha \beta(h_\alpha)=0
\eeq    
where the $I$ runs over colored faces and $\alpha$ runs over nodes that are fully
contained in the shaded region (i.e. nodes such that all the edges connected to 
them are on the boundary of at least one shaded face). The value of $c_\alpha$ is 2 
if all the faces around the given node are colored and 1 otherwise. This idea is a 
generalization of the one introduced in \cite{Hanany:1998ru} for brane boxes.  

\begin{figure}[ht]
  \epsfxsize = 10cm
  \centerline{\epsfbox{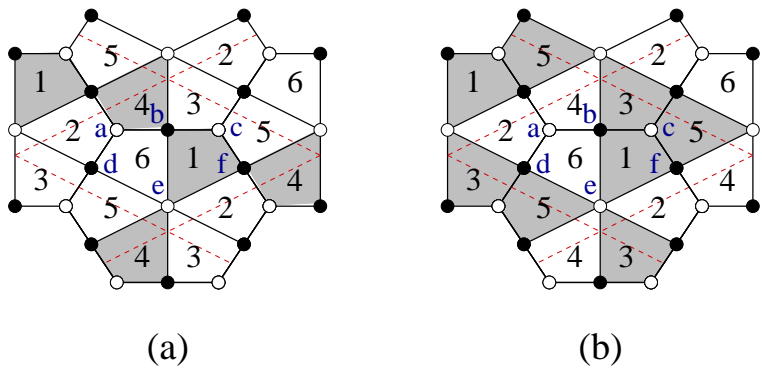}}
  \caption{\small Tilings corresponding to deformation branes of model I of $dP_3$. a) $(1,0,0,1,0,0)$ brane. 
b) $(1,0,1,0,1,0)$ brane.}
  \label{dimers_betas1}
\end{figure}

Lets us now see how these ideas apply to model I of $dP_3$. \fref{dimers_betas1} 
shows the tilings
for the $(1,0,0,1,0,0)$ and $(1,0,1,0,1,0)$ fractional branes. We have indicated the fundamental cell
and also labeled nodes.

From \fref{dimers_betas1}.a, we see that the $(1,0,0,1,0,0)$ brane gives rise to 
the following relation between the beta functions for $g_1$ and $g_4$ and the ones 
for the couplings $h_b$ and $h_f$ of two quartic superpotential terms. 

\beq
\beta_{g_1}+\beta_{g_4}-\beta_{h_b}-\beta_{h_f}=0
\eeq

From \fref{dimers_betas1}.b, we deduce the following relation between the beta 
functions for the gauge couplings $g_1$, $g_3$, $g_5$, and the 
superpotential couplings $h_c$ (cubic term) and $h_e$ (sixth order term).

\beq
\beta_{g_1}+\beta_{g_3}+\beta_{g_5}-2 \beta_{h_c}-\beta_{h_e}=0
\eeq

\section{Conclusions}
\label{conclusions}

In this paper we have studied the gauge theory dynamics on systems of fractional 
branes (possibly in the presence of D3-branes) at toric CY singularities. 
This has lead to a classification of fractional branes in terms of the IR behavior of the 
gauge theory. We have recovered the known cases where this dynamics preserves supersymmetry, and leads to 
either confinement or $N=2$ dynamics. Moreover, we have observed that the generic fractional 
brane leads to breakdown of supersymmetry due to non-perturbative superpotentials. When the dynamics of 
FI terms (related to localized 
closed string moduli) or baryonic operators is taken into account, we are lead to runaway behaviors as discussed in section \ref{section_dynamical_FI}. Geometrically, non-supersymmetric 
behavior occurs whenever the fractional brane does not have an associated complex 
deformation. For toric geometries, we have studied how complex deformations are 
easily characterized in terms of possible splittings of the web diagram into subwebs 
in equilibrium. This criterion translates into a rigorous prescription in terms
of possible decompositions of the toric polytope into Minkowsky sum of sub-polytopes
(see appendix \ref{obstruction}).
 
Our results have important implications concerning the dual supergravity throats 
describing the UV duality cascades of such gauge theories. In particular, they show 
that the naked singularities of such throats generically cannot be smoothed by the 
standard mechanism of complex deformation (since, in the best of cases, such deformations exist at a given 
order but are obstructed at higher orders being impossible to integrate them to full deformations). 
The smoothing should rather be non-supersymmetric (hence going far beyond the standard ansatz of conformal CY 
metrics with ISD 3-form fluxes), and should very possibly lead to unstable backgrounds. 
Hence, the 
expectation from the gauge theory side is that there exists no smooth supergravity 
configuration obeying the equations of motion, and with the asymptotics of the 
warped throats with 3-form fluxes (for choices of fluxes corresponding to DSB 
fractional branes).

It would be very interesting to find independent information on the 
nature of the singularities for such supergravity throats, in order to find more 
direct information about them. One possible tool \footnote{We thank D. Martelli for 
discussions on this point.} would be the introduction of e.g. D5-branes wrapped on 
the 2-cycle of $Y^{p,q}$ as a probe of the infrared limit of the throats in 
\cite{Ejaz:2004tr}.

We have also shown that brane tiling configurations are useful tools for the study 
of deformation and $N=2$ fractional branes. It would be nice to develop an 
appropriate picture of DSB branes, and its non-perturbative dynamics, in this 
language.

We expect much progress in addressing these and other questions in the setup of branes 
at singularities and its dual versions.

\section*{Acknowledgements}

We thank J.F.G. Cascales, Y. H. He, S. Kuperstein, M. Mahato, D. Martelli, C. Nu\~nez, 
Y. Oz, J. Sonnenschein, L. Pando Zayas, R. Russo and D. Vaman for useful 
discussions. F. S. thanks the CERN PH-TH Division for hospitality during completion 
of this work and the spanish Ministerio de Educacion y Ciencia for financial support through an F.P.U grant.
A. M. U. thanks M. Gonz\'alez for encouragement and support.
The research of S. F. and A. H. was supported in part by the CTP and LNS of
MIT and the U.S. Department of Energy under cooperative research
agreement $\#$ DE--FC02--
94ER40818, and by BSF American--Israeli Bi--National Science Foundation. A. H. is 
also indebted to a DOE OJI Award. The research of A. M. U. and F. S. was supported 
by the CICYT, Spain, under project FPA2003-02877, and by the networks
MRTN-CT-2004-005104 `Constituents, Fundamental Forces and Symmetries of
the Universe', and MRTN-CT-2004-503369 `Quest for Unification'.

\appendix
\section{Deformations and obstructions for isolated toric singularities}
\label{obstruction}

In this appendix we describe some useful results by K. Altmann
\cite{altmann1,altmann2} (see \cite{altmann3} for a complete discussion),
concerning the possible complex deformations of toric singularities, and
their obstructions.

Three-dimensional Gorenstein (i.e. CY) toric singularities are 
described by toric cones with base given by a polytope (compact polygon)
lying on a 2-plane. Let us denote $a^i$, $i=1,\ldots, N$ the 2d vectors
defining the positions of its vertices, and $d^i=a^{i+1}-a^i$ the edges
(with the understanding that $d^N=a^1-a^N$). In order to be isolated, the
polytope should have primitive edges. Namely there are no points on the
edges of the polygon, i.e. $d^{i+1}\neq d^i$ for all $i$.   
   
Given two 2d polygons $P_1$, $P_2$, one can define its Minkowski sum as
the polygon $P=P_1+P_2$ defined by $P = \{ p=p_1+p_2\, |\, p_1\in P_1, p_2\in 
P_2 \}$. Namely the set of vectors given as sums of vectors in the summand
polygons. It is easy to realize that the polygon $P$ has edges given by
the union of the sets of edges of the polygons $P_1$, $P_2$.

\begin{figure}[ht]
  \epsfxsize = 10cm
  \centerline{\epsfbox{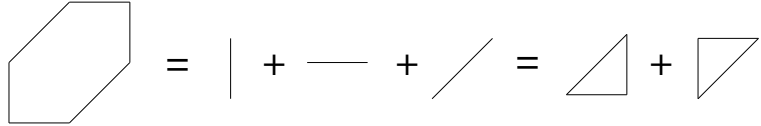}}
  \caption{The two decompositions of the toric polygon of the cone over $dP_3$
as a Minkowski sum. The two decompositions correspond to the two
splittings into subwebs in figure \protect\ref{dp3}.
}
  \label{minkowski}
\end{figure}

The complex deformations of isolated Gorenstein singularities are
completely characterized in terms of the possible decompositions
of the toric polytope into a Minkowski sum of polytopes. It is easy to see
that this is equivalent to the criterion used along the paper that the web
diagram (dual to the toric polygon) splits into a set of subwebs in
equilibrium (dual to the summand polygons). Figure \ref{minkowski} shows
the two decompositions of the polytope for the complex cone over $dP_3$,
whose relation to the dual picture of splitting into subwebs (figure   
\ref{dp3}) is manifest.

Following \cite{altmann1,altmann2}, the parameter space of complex
deformations of a isolated CY singularity, allowed up to order $K$
(i.e. not obstructed up to order $K+1$) is the subspace of $\IC^N$ obeying
the equations
\beqa
\sum_{i=1}^N (t_i)^k \, d^i \, = \, 0   \quad \quad 0<k\le K
\eeqa
Since the equations are homogeneous, there is one overall scaling which
should be removed, leading to a projective parameter space.
True complex deformations satisfy the above equation for all $k>0$. In  
fact, it is enough to verify this up to a (polytope dependent) finite value $K_0$, 
which guarantees the equations for $k>K_0$ are also satisfied \footnote{The value 
$K_0$ can be obtained as follows: define the polytope as the intersection set of a 
strips in the 2d space. $K_0$ is the lattice thickness of thickest such strip.}.

For true complex deformations, solutions $t_i$ to the above equations parametrize 
the base space of the so-called versal deformation space. This means that any
complex deformations of the defining equations of the toric singularity
can be written in terms of the $t_i$'s, which are hence the deformation
parameters. A complex deformation can be associated to a Minkowski
decomposition of the polytope, by gathering edges with same value of $t_i$ in the 
solution.

This description provides a characterization of true complex deformations,
and moreover of deformations valid at order $K$ but obstructed at order 
$K+1$. Let us apply this formalism to some situations of interest.

For any toric singularity of the kind described above, the first order
deformations lie in a subset of $\IC^N$ given by
\beqa
\sum_i t_i \, d^i \, =\, 0
\eeqa
These are two equations for $N$ variables, and we should remove the
scaling. This leaves $N-3$ first order deformations. This shows that
the cone over $dP_0$ has no deformations, and e.g. that the cones over
$Y^{p,q}$ have one first order deformation. 

Let us consider the $Y^{p,q}$ theories, whose toric diagram is shown in
figure \ref{ypq}, in more detail.
Since for $q\neq 0$ no splitting of the web diagram into subwebs in
equilibrium, we expect that the first order deformations are obstructed at
some higher order, for $q\neq 0$. This can be verified explicitly with
our above formula. We have $a^1=(0,0)$, $a^2=(1,0)$, $a^3=(0,p)$,
$a^4=(-1,p-q)$, and hence $d^1=(1,0)$, $d^2=(-1,p)$, $d^3=(-1,-q)$,
$d^4=(1,-(p-q))$.

The equations for the space of complex deformations read
\beqa
t_1^{\, k}\, -\, t_2^{\, k}\, -\, t_3^{\, k}\, +\, t_4^{\, k}\, = \, 0 \nonumber \\
p\, t_2^{\, k}\, -\, q\, t_3^{\, k}\, -\,(p-q)\, t_4^{\, k}\, = \, 0
\eeqa
The case $q=0$ stands out as special. The equations correspond to
\beqa
t_1^{\, k}\, -\, t_2^{\, k}\, -\, t_3^{\, k}\, +\, t_4^{\, k}\, = \, 0 \nonumber \\
p\, t_2^{\, k}\, -\, p \, t_4^{\, k}\, = \, 0
\eeqa
At linear order we have
\beqa
t_1\, -\, t_2 \, -\, t_3 \, +\, t_4\, = \, 0 \nonumber \\
t_2\, - \, t_4\, = \, 0
\eeqa
Hence $t_4=t_2$ and $t_1=t_3$. So there is one complex deformation. It is
easy to see that it obeys all the equations for $k>1$, so it is not
obstructed at any order. This is in agreement with the fact that
$Y^{p,0}$ is an orbifold of the conifold, with the quotient preserving
the complex deformation of the conifold. Notice that the absence of 
obstruction is nicely linked to the relations between the $d^i$'s, and 
hence to the possibility of web recombination. Concretely, the complex
deformation can be described as the splitting into two subwebs given by
straight lines, or in terms of a decomposition of the polytope as a Minkowski sum of 
two segments.

Let us consider the generic case $q\neq 0$. For $0<q<p$, we have at linear
order
\beqa
t_1\, -\, t_2\, -\, t_3\, +\, t_4\, = \, 0 \nonumber \\
p\, t_2\, -\, q\, t_3\, -\,(p-q)\, t_4\, = \, 0
\label{rel1}
\eeqa
This gives
\beqa
t_1\, =\, \frac{p+q}{p} t_3 \, - \, \frac{q}{p} t_4 \quad , \quad
t_2\, =\, \frac{q}{p} t_3 \, + \, \frac{p-q}{p} t_4
\eeqa
Hence there is a two-parameter solution to the equations, one of which is
complete rescaling. Hence there is one deformation at first order.

At second order, we have to impose the additional conditions
\beqa
t_1^{\, 2}\, -\, t_2^{\, 2}\, -\, t_3^{\, 2}\, +\, t_4^{\, 2}\, = \, 0
\nonumber \\
p\, t_2^{\, 2}\, -\, q\, t_3^{\, 2}\, -\,(p-q)\, t_4^{\, 2}\, = \, 0
\eeqa
Using (\ref{rel1}) in the first equation, one obtains
\beqa
q(t_3-t_4)^2=0
\eeqa
hence for $q\neq 0$ we obtain $t_3=t_4$. Using the rest of the equations
one easily obtains $t_1=t_2=t_3=t_4$. This simply corresponds to a total
rescaling and does not describe a complex deformation. Hence there is no
complex deformation at second order, namely the first order complex
deformation is obstructed at second order.

This mathematical result explains that, despite the first order deformation, 
constructed explicitly in \cite{Burrington:2005zd}, the geometries $Y^{p,q}$ do
not admit complex deformations, and hence they cannot provide a smoothing
of the naked singularities in the throats in \cite{Ejaz:2004tr}.

It is straightforward to describe in this language complex deformations
mentioned in the main text in terms of splitting into subwebs in
equilibrium. An interesting point is that for non-isolated Gorenstein
singularities the description of complex deformations in terms of
Minkowski sums is not complete. Physically, one can clearly associate
these new mathematical subtleties to the appearance of a different kind of
fractional branes (namely $N=2$ fractional branes) for non-isolated toric
singularities.


\bibliographystyle{JHEP}

\end{document}